\documentclass[aps,prapplied,showpacs,twocolumn,superscriptaddress,10pt]{revtex4-2}
\usepackage{amsmath}
\usepackage{amsfonts}
\usepackage{braket}
\usepackage{amssymb}
\usepackage{graphicx}
\usepackage[colorlinks=true,linkcolor=blue,urlcolor=black,citecolor=blue]{hyperref}
\usepackage{comment}

\newcommand{\da}{\dagger}

\newcommand{\diag}{\mathop{\mathrm{diag}}}

\begin{document}
\title{Tunable Tradeoff between Quantum and Classical Computation via Nonunitary Zeno-like Dynamics}

\author{P. V. Pyshkin}
\thanks{Corresponding author. pavel.pyshkin@gmail.com}
\affiliation{Department of Physical Chemistry, The University of the Basque Country UPV/EHU, 48080 Bilbao, Spain}
\affiliation{Department of Physics, University of the Basque Country UPV/EHU, 48080 Bilbao, Spain}
\affiliation{Institute for Solid State Physics and Optics, Wigner Research Centre, P.O. Box 49, H-1525 Budapest, Hungary}
\author{A. G\'abris}
\affiliation{Czech Technical University in Prague, Faculty of Nuclear Sciences and Physical Engineering B\v rehov\'a 7, 115 19 Praha 1, Star\'e M\v esto, Czech Republic.}
\affiliation{Institute for Solid State Physics and Optics, Wigner Research Centre, P.O. Box 49, H-1525 Budapest, Hungary}

\author{Da-Wei Luo}
\affiliation{Center for Quantum Science and Engineering and Department of Physics,Stevens Institute of Technology, Hoboken, New Jersey 07030, USA}

\author{J. Q. You}
\affiliation{Department of Physics and State Key Laboratory of Modern Optical Instrumentation, Zhejiang University, Hangzhou 310027, China}

\author{Lian-Ao Wu}
\thanks{lianaowu@gmail.com}
\affiliation{Department of Physics, University of the Basque Country UPV/EHU, 48080 Bilbao, Spain}
\affiliation{Ikerbasque, Basque Foundation for Science, 48011 Bilbao, Spain}
\affiliation{EHU Quantum Center, University of the Basque Country UPV/EHU, Leioa, Biscay 48940, Spain}

\date{\today}
\begin{abstract}
    We propose and analyze a nonunitary variant of the continuous time Grover search algorithm based on frequent Zeno-type measurements. We show that the algorithm scales similarly to the pure quantum version by deriving tight analytical lower bounds on its efficiency for arbitrary database sizes and measurement parameters. We also study the behavior of the algorithm subject to noise, and find that under certain oracle and operational errors our measurement-based algorithm outperforms the standard algorithm, showing robustness against these noises. Our analysis is based on deriving a non-hermitian effective description of the algorithm, which yields a deeper insight into components responsible for the quantum and the classical operation of the protocol. 
\end{abstract}

\maketitle

\section{Introduction}

Quantum measurement has been proven to be a powerful tool that not only allows us to learn about a quantum system but also to control its state.
It plays a fundamental role in quantum information with applications, among many other things, ranging from quantum communications to quantum algorithms.
The quantum Zeno effect (QZE) is a widely-employed technique for quantum control, which is based on repeated frequent measurements of the entire system or part of it \cite{Facchi_2008}.
A number of studies have employed the QZE or similar techniques for various flavours of search problems \cite{rudolph_quantum_2002, Farhi-search-by-measurement, hosten_counterfactual_2006}, to establish remarkable relations between the efficiency of quantum and associated classical algorithms \cite{bomb-complexity-lin_et_al}, as well as singular value transformations \cite{gilyen_quantum_2019}.
The Zeno dynamics of a closed quantum system induced by projective measurement will yield unitary dynamics, but the evolution due to observation may be more general.
Measurement-induced nonunitary dynamics have been considered in the literature both as primitives \cite{knill_scheme_2001,non-unitary-gates-2005,non-unitary-QW-1-PhysRevA.71.022307,non-unitary-QW-2-KENDON2008187,non-unitary-compute-1-PhysRevA.96.032321}, or as essential ingredients of specific protocols \cite{Hiromichi, Wu_entanglement_generation, Li2011,gsc_paper, Zeno-Polaron-PRB, Filippov2017, Streltsov2011_entanglment_by_measurements, Gilyen, Piani2014_entanglment_by_measurements, Torres2017,Pyshkin2017, two-qubit-entangl-measurements-non-herm}.
Another source of nonunitary evolution may be a special coupling with environment~\cite{Amin2008, Luo2015, Novo2018}. 

In this paper, we consider an algorithm that is a variant of the continuous search algorithm introduced by Farhi and 
Gutmann~\cite{H-oracle-first-Farhi1998}.
This algorithm follows a scheme based on the combination of time-dependent measurement and Hamiltonian evolution of the system \cite{Hiromichi}, admitting a nonunitary description and exhibiting a non-periodic time dependence of the target fidelity.
Our approach is based on repeated measurements and post-selection, therefore the survival probability associated with successfully completing the desired number of steps may be less than one, in addition to the usual probability related to the target fidelity.
We show that in the case of a detuned oracle, the target fidelity can be increased up to unity at the expense of the survival probability, which makes it a favourable choice in situations where the correctness of the obtained result cannot be easily verified. While the algorithm is interesting in its own right, it is remarkable that our measurement-based algorithm is robust and self-protected against a certain class of noises~\cite{self-protected-algos,PRA-recoupling-and-decoupling}.
This robustness can be attributed to the noise suppression of the repeated measurements.
The algorithm provides a framework for studying the trade-off between quantum and classical computation, where the quantum speedup is related to the unitary operations while measurements lead to the appearance of classical probabilities for different outcomes.
Thus, in order to combine classical and quantum computations in a single process, one can consider unitary dynamics interrupted by selective measurements.

\section{The algorithm}

\begin{figure*}
    \begin{center}
        \includegraphics{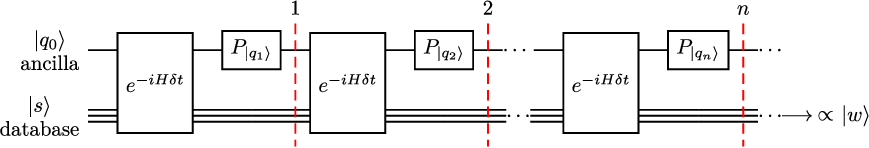}
    \end{center}
    \caption{Circuit diagram of the nonunitary algorithm. The ancilla states are $\ket{q_n}= \cos\theta_n \ket{\uparrow} + \sin\theta_n\ket{\downarrow}$ with $\theta_n=\theta_0+n\delta\theta$ as defined in the text. The operation $P_{\ket{q_n}}$ denotes the projection onto the state $\ket{q_n}$ as a result of the measurement and post-selection. As shown in Sec.~\ref{sec:nonunitary-desc}, the state of the database register asymptotically approaches the marked element $\ket{w}$ up to a factor reflecting the survival probability.}
    \label{fig:circuit-diagram}
\end{figure*}

We start from the continuous-time Grover search algorithm described by the Hamiltonian in the Hilbert space~$\mathcal{H}_N$ ($\dim\mathcal{H}_N = N$),
\begin{equation}\label{contin_grover}
    H_G = H_o + H_d = -\ket{w}\bra{w} - \ket{s}\bra{s},
\end{equation}
where $\ket{s}=\sum_n^N \ket{n}/\sqrt{N}$ and $\ket{w}$ denotes the marked element in the database.
Taking $\ket{\psi(0)} = \ket{s}$ as the initial state the evolution remains in a two-dimensional subspace of the total Hilbert space spanned by the basis vectors~$\{ \ket{w}, \ket{r}  \}$, where $\ket{r} = (\ket{s} - x\ket{w})/\sqrt{1-x^2}$, and $x = 1/\sqrt{N}$.
The evolution of the initial state is given by~\cite{H-oracle-first-Farhi1998} 
\begin{multline}\label{ideal_grover_evolution}
    \ket{\psi(t)} = e^{-iH_Gt}\ket{s} = e^{it}\left\{ \vphantom{\sqrt{1-x^2}}\left(\vphantom{1^1}x\cos(xt) + i \sin(xt)\right)\ket{w} \right.\\
    + \left.\sqrt{1-x^2}\cos(xt)\ket{r}   \right\},
\end{multline}
where we have taken~$\hbar = 1$ for convenience.
As follows from~(\ref{ideal_grover_evolution}) the probability to have the system in the target state~$\ket{w}$ oscillates in time, with maxima at~$T_j = \pi x^{-1} (1/2 + j) \propto\sqrt{N}$, where $j=0,1,2,\dots$. 
This periodic behavior is a consequence of the unitarity of the evolution. 
We stress here that issue of this periodic behaviour can also be tacked by adiabatic quantum search algorithms \cite{Adiabatic_qc_1,Adiabatic_qc_2}, as well as sophisticated time-dependent protocols \cite{non-adiab-search-1, non-adiab-search-2}.

It is well known that the nonunitary dynamics of quantum systems can have an asymptotic steady state instead of non-damping oscillations (see e.g.\  Ref.~\cite{Wu_entanglement_generation, Li2011, Filippov2017,Zeno-Polaron-PRB}). 
We have designed a modification of this algorithm so that $\ket{w}$ becomes such a steady-state.

Let us introduce our nonunitary protocol. 
We add a qubit ancilla to our system and extend its Hamiltonian to
\begin{equation}\label{ancila-H}
H = -\mathcal{I}\otimes\ket{w}\bra{w} - \sigma_z\otimes\ket{s}\bra{s},
\end{equation}
with $\mathcal{I}$ and $\sigma_z$ being the identity and Z-Pauli matrices acting in a space~$\mathcal{H}_A = \{\ket{\uparrow}, \ket{\downarrow}\}$ of the ancilla qubit.
Thus the joint Hilbert space now is $\mathcal{H}_A\otimes\mathcal{H}_N$. 
We underline here that the interaction between oracle and ancilla ``does not know'' about~$\ket{w}$ state.
   
We consider the continuous evolution interrupted by projective measurements.
The $n$-th step of the protocol of nonunitary evolution is the following:
\begin{enumerate}\itemsep1pt
\item[1)] The initial state of joint system is $\ket{q_{n-1}}\otimes\ket{\psi_{n-1}}$, where ancilla state $\ket{q_{n-1}} = \cos\theta_{n-1} \ket{\uparrow} + \sin\theta_{n-1}\ket{\downarrow}$
\item[2)] The joint system evolves time~$\delta t$ driven by Hamiltonian~(\ref{ancila-H}). It is worth noting that $\delta t$ is not necessarily small, but we assume $\delta t\ll T_0$.
\item[3)] One performs projection measurement~$\ket{q_n}\bra{q_n}$ on the ancilla, where $\ket{q_{n}} = \cos(\theta_{n-1}+\delta\theta) \ket{\uparrow} + \sin(\theta_{n-1}+\delta\theta)\ket{\downarrow}$, where~$\delta\theta = \alpha x \delta t \ll1$, with $\alpha$ being a tunable parameter.
\item[4)] A successful outcome occurs with probability $p_n$, and one leaves the state of joint system in the state $\ket{q_n}\otimes\ket{\psi_n}$. Otherwise, an unsuccessful outcome indicates that this run of the algorithm must be aborted.
\item[5)] Let $\theta_{n} = \theta_{n-1} + \delta\theta$
\end{enumerate}
See Fig.~\ref{fig:circuit-diagram} for a circuit diagram representation of the complete algorithm.
We would like to point out that the algorithm does not require reinitialization of the ancilla qubit after each measurement to some initial state, and this can be considered an advantage.

\begin{figure}[b]
    \begin{center}
        \includegraphics{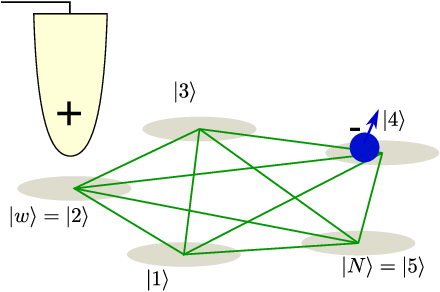}
    \end{center}
    \caption{Possible proof of concept realization of our proposed algorithm, with spin-dependent hopping of spin-1/2 particle on a complete graph. The marked node~$\ket{w}$ has an additional potential (electrostatic gate). The spin degree of freedom plays the role of an ancilla qubit.}
    \label{fighopping}
\end{figure}

The Hamiltonian in Eq.~(\ref{ancila-H}) can describe single charged spin-$1/2$ particle on a complete graph~\cite{Complete-graph-hopping-Salerno1995} with unconstrained and spin-dependent hopping~\cite{spin-dependent-hopping-PhysRevB.95.155120}, with $w$ corresponding to some unknown vertex where the electrostatic gate is applied (see Fig.~\ref{fighopping}). In such case, Eq.~(\ref{ancila-H}) can be rewritten as $H = -c^\da_{w\uparrow} c_{w\uparrow} - c^\da_{w\downarrow} c_{w\downarrow} - \sum_{i\neq j} (c^\da_{i\uparrow}c_{j\uparrow} - c^\da_{i\downarrow}c_{j\downarrow}) $, where $c_{i\uparrow(\downarrow)}$ is annihilation operator for the particle on site~$i$ with spin~$\uparrow(\downarrow)$. In this setting, the spin degree of freedom acts as ancilla.

\begin{figure} 
    \centering
    \includegraphics{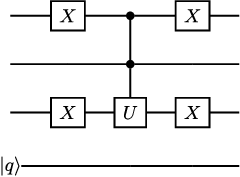}
    
    \caption{Realization of the operator $\exp(i\ket{w}\bra{w}\delta t/m)$ via the circuit model for $N = 2^3$ and $\ket{w}=\ket{010}$. Controlled unitary is~$U = \diag(1, e^{ i \delta t/m})$. The bottom wire corresponds to the ancilla qubit which is not involved in the construction of the oracle.}
    \label{fig-circuit-oracle}
\end{figure}

\begin{figure} 
    \centering
    \includegraphics{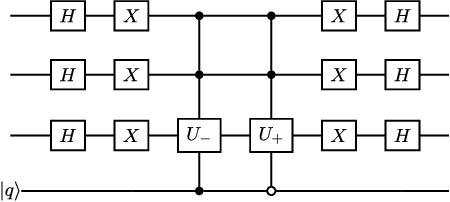}
    
    \caption{Realization of the operator $\exp(i\sigma_z\otimes\ket{s}\bra{s}\delta t/m)$ via the circuit model for $N = 2^3$. Controlled unitaries are~$U_{\pm} = \diag(1, e^{\pm i \delta t/m})$. The bottom wire corresponds to the ancilla qubit.}
    \label{fig-circuit}
\end{figure}

Another way to implement the Hamiltonian~(\ref{ancila-H}) (at least as a proof of concept) with the common quantum circuit model is to use a Trotterization technique~\cite{Trotterization-Lloyd1996,Wu2002-trotterization,Trotterization-Smith2019,Trotterization-Tacchino2020}. In this approach we can use the Trotter formula
\begin{equation}\label{trotter}
e^{-iH\delta t} = \lim_{m\rightarrow\infty}\left( e^{i\ket{w}\bra{w}\delta t/m} e^{i\sigma_z\ket{s}\bra{s}\delta t/m} \right)^m.
\end{equation}
For real simulations, one should use some finite value of~$m$ in~(\ref{trotter}). In order to implement the unitary operation for the first multiplier in the round brackets (oracle) in~(\ref{trotter}) one can use the algorithm depicted in Fig.~\ref{fig-circuit-oracle}, while for the second one (diffusion operator) one can use the algorithm depicted in Fig.~\ref{fig-circuit}.

\section{nonunitary description}
\label{sec:nonunitary-desc}

The stroboscopic dynamics described above corresponds to a transformation of the initial state $\ket{s}$ by a nonunitary operator~$V(n)$ via
\begin{equation}\label{result_non-u-general}
    \ket{\psi_n} = \frac{V(n)\ket{s}}{\sqrt{\braket{s|V^\dagger (n) V(n)|s}}},
\end{equation}
where
\begin{equation}\label{V-general}
    V(n) = \prod_{j=1}^{n}V_j, \mbox{ and } V_j = \braket{q_{j}|e^{-iH\delta t}|q_{j-1}}.
\end{equation}
Note, that the terms in the above product are ordered from right to left. 
The survival probability~$P(n)$ of $n$ first steps is hence given by $P(n)=p_1p_2\dots p_n=\braket{s|V(n)^\dagger V(n)|s}$.
Equations~(\ref{result_non-u-general}) and~(\ref{V-general}) can be considered as a set of POVM measurements in the Hilbert space of the oracle. One can add operators~$V'_j=\braket{q^\perp_j|e^{-iH\delta t}|q_{j-1}}$ with $\ket{q^\perp_j}=\sin(\theta_j)\ket{\uparrow} - \cos(\theta_j)\ket{\downarrow}$ to have a full set of POVM operators for each step~$\{V_j^\dagger V_j, V'^\dagger_j V'_j\}$. It is easy to check $V_j^\dagger V_j + V'^\dagger_j V'_j = \mathbb{I}$.
We define the target fidelity at step~$n$ as $f(n) = |\braket{w|\psi_n}|^2$. 
Together with survival probability~$P(n)$ these are the main characteristics of our nonunitary process.

Exact analytic calculation of~$V(n)$ in~(\ref{V-general}) is difficult, however, at first we can note that Hamiltonian~(\ref{ancila-H}) has a block-diagonal structure and the calculation of a single~$\delta t$ step is an easy task. 
Taking $\theta_0=0$ we obtain the expression
\begin{equation}
V(n) = \prod_{j=1}^{n} \left( C_je^{-i(H_o+H_d)\delta t} + S_je^{-i(H_o-H_d)\delta t} \right),
\end{equation}
with $C_j = \cos((j-1)\delta\theta)\cos(j\delta\theta)$ and $S_j = \sin((j-1)\delta\theta)\sin(j\delta\theta)$.

Assuming $x\rightarrow0$ and $x\delta t\ll 1$ (but finite $\delta t$) we can write, up to a global phase,
\begin{equation}\label{V_j-1}
    V_j = \begin{pmatrix}
    C_j+S_j       &  iC_jx\delta t - \frac{S_jx}{2}(1 - e^{-2i\delta t})   \\
    iC_jx\delta t - \frac{S_jx}{2}(1 - e^{-2i\delta t})       & C_j + S_je^{-2i\delta t} \\
    \end{pmatrix}.
\end{equation}
Further we consider $\delta\theta\ll1$.
There are special values of~$\delta t$ corresponding to $\exp(-2i\delta t) = 1$
\begin{equation}\label{t-pi-k}
    \delta t = \pi k, \quad k=1,2,3,\dots,
\end{equation}
when each term $V_j$ can be approximated by a rescaled unitary, hence the same will hold for the whole process $V(n)$.

\begin{figure}[b]
    \begin{center}
        \includegraphics{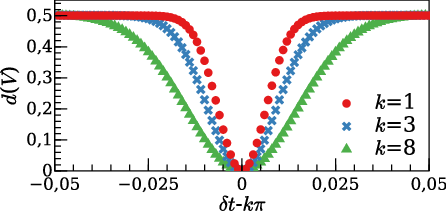}
    \end{center}
    \caption{Distance from unitarity for a complete process as a function of~$\delta t$.}
    \label{figdistance}
\end{figure}

As a measure of (non-)unitarity we use the distance 
\begin{equation}
d(V) = 1 - \frac{1}{2}\mathrm{Tr}(V^\dagger V)
\end{equation}
employing the Frobenius norm.
We can use this measure to track the trade-off between the unitarity (i.e. quantum computation) and non-unitarity (i.e. classical computation) of the process.

To illustrate the analytical result (\ref{t-pi-k}) we plot in Fig.~\ref{figdistance} the dependence of $d(V)$ of the resulting transformation~$V(n_G)$ (with $n_G = \lfloor \pi \sqrt{N}/ (2\delta t)\rfloor$) for various values of $\delta t$ with $N=10^{10}$ and $\delta \theta / \delta t = 3\times10^{-6}$.
The minima can be determined from Eq.~(\ref{t-pi-k}), which involves neither the parameter~$\delta\theta$, nor the index~$j$. 

For further analysis let us define a non-hermitian effective Hamiltonian $H_{\mathrm{eff}}(t)$ by taking the formula
\begin{equation}\label{Heff-def}    
V_j = e^{-iH_{\mathrm{eff}}(j\delta t)\delta t}
\end{equation} 
for each $j$, and extending $H_{\mathrm{eff}}(t)$ as piecewise constant on the intervals $( (j-1)\delta t, j\delta t]$.
The resulting time-dependent (non-hermitian) Hamiltonian provides an equivalent description of the search algorithm.

\begin{figure}
    \begin{center}
        \includegraphics{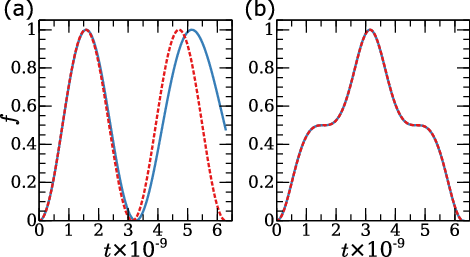}
    \end{center}
    \caption{Fidelity of matching target state versus the time with $N=10^{18}$, $\delta t = 10^6\pi$ and $\delta\theta/\delta t = 0.1x$ for solid line in panel (a) and $\delta\theta/\delta t = x$ for solid line in panel (b). Dashed line in panel (a) corresponds to continuous Grover dynamics~(\ref{ideal_grover_evolution}). The dashed line in panel (b) is calculated using the approximation (\ref{V-unitary}), which shows perfect agreement with the direct simulation.}
    \label{figunitary}
\end{figure}

Introducing the parameter $\tau$ which describes the level of non-unitarity as $\delta t = \pi k + \tau$, for some integer $k$, and assuming $|\tau| \ll 1$ we can approximate the piecewise constant non-hermitian Hamiltonian by the continuous expression
\begin{multline}\label{H-eff}
H_{\mathrm{eff}}(t)\approx - x \cos^2(\alpha x t) \left(\vphantom{1^1}\ket{w}\bra{r}+\ket{r}\bra{w}\right) + \\
2\frac{\tau}{\delta t}\sin^2(\alpha x t)\ket{r}\bra{r} -2 i\sin^2(\alpha x t)\frac{\tau^2}{\delta t}\ket{r}\bra{r}, 
\end{multline}  
where $\alpha x= \delta\theta / \delta t$. We note that when $\tau=0$ the right hand side becomes hermitian, thus the formal expression $V(n)\approx \mathcal{T}\exp(-i\int_0^{n\delta t} H_{\mathrm{eff}}(t')dt')$ permits an approximation by purely unitary dynamics, 
\begin{multline}\label{V-unitary}
V(n)\approx \mathbb{I}\cos(A(n)) + i\left(\vphantom{1^1}\ket{w}\bra{r}+\ket{r}\bra{w}\right)\sin(A(n)), \\
 A(n) = \frac{x\delta t}{2}\left( n + \frac{\sin(2n\delta\theta)}{2\delta\theta}  \right).
\end{multline}
Note, that in the limit~$\delta\theta\rightarrow0$ we have~$A(n) = xn\delta t$, yielding the standard continuous Grover algorithm with $f(n) = \sin^2(xn\delta t)$. The first maximum occurs at time $T_0 = n_G\delta t = \pi\sqrt{N}/2$, while 
for $\delta\theta / \delta t = x$ the first maximum is reached at time $\pi\sqrt{N}$. In Fig.~\ref{figunitary} we plot some numerical examples.

At non-zero values of $\tau$ we expect the non-hermitian component to drive the system to a steady-state.
Of particular interest is the emergence of an asymptotic dynamics consisting of the target state $\ket{w}$ alone, manifesting in the saturation of the target fidelity at the value of~$1$.
A heuristic analysis of the Hamiltonian~(\ref{H-eff}) shows that in order to achieve saturation dynamics of the fidelity, the relation  $\sqrt{x\delta t}\ll \tau \ll 1$  must hold between control parameters (see Appendix for details).

It's worth pointing out that in our algorithm, when we choose~$\theta_0 = \pi/4$ and $\delta t = \pi/2 + \pi k$, $k = 0,1,2,\dots$, the first iteration of our algorithm may be approximately described via $V_1\approx \ket{w}\bra{w}$ (up to corrections of order $x\delta t \ll 1$). In this situation we would have a {\em classical search regime}, i.e. we can organize {\em projection onto the unknown state}~$\ket{w}$, and the probability of a successful outcome of the ancilla measurement for initial state~$\ket{s}$ is $\braket{s|V_1^\da V_1|s}\approx 1/N$, which corresponds to the efficiency of the classical search~$O(N)$. This example demonstrates the possibility of a classical computation regime which is the opposite of the quantum limit discussed after Eq.~(\ref{V-unitary}). Thus we have shown that the proposed algorithm contains quantum, classical and intermediate regimes depending on the parameters chosen.

\section{Scaling properties}
To study the scaling properties of our nonunitary dynamics we investigate the following transformation of the Hamiltonian~(\ref{H-eff}): $xt = t'$, $H'(t')= H(t)/x$  while keeping $x\delta t$ and $\tau$ constant, and assuming  $x\delta t \ll 1$.
The last condition here can be approximately satisfied by properly choosing both~$N\gg1$ and an integer~$k$. 
If we have~$V(n)$ for some $N=N_1$ and $\delta t_1 = k_1\pi + \tau$, then we can find $N_2>N_1$ and $\delta t_2 = k_2\pi + \tau$ which corresponds to approximately the same operator~$V(n)$ when~$k_2$ is an integer such that
\begin{equation}\label{k2}
k_2 \approx \frac{1}{\pi}\sqrt{\frac{N_2}{N_1}}(k_1\pi + \tau) - \frac{\tau}{\pi}.
\end{equation}
This means that we have almost the same number of steps of the protocol for different~$N$, and the duration~$\delta t$ of each step is proportional to~$\sqrt{N}$. 
As can be seen, the needed relative accuracy of timing~$\tau / \delta t \propto 1/\sqrt{N}$ grows with~$N$.

 \begin{figure}
    \begin{center}
        \includegraphics{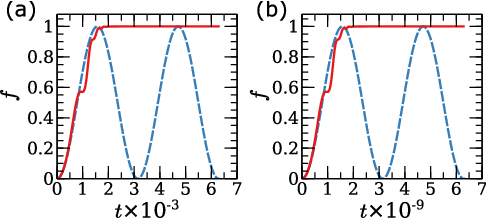}
    \end{center}
    \caption{Illustration of the scalability of the proposed algorithm. We show the fidelity of matching target state versus the time for common Grover algorithm (dotted line) and nonunitary search (solid lines) with~$\alpha=0.3$ and~$\tau=0.2$. (a) $N=10^{6}$, $k=1$ (b) $N=1.000000162505052417\times10^{18}$, $k=1.063662\times10^6$.}
    \label{fig-scaling}
\end{figure}

The recurrence relation Eq.~(\ref{k2}) can be used to characterize the scaling properties of the algorithm.
Assume that we have a certain process with fidelity~$f_1(t)$, survival probability~$P_1(t)$, and this process is characterized by~$N_1$, $\delta t_1$, and $\delta\theta_1$. 
We argue that for any requested database size~$N_r\gg N_1$ we can find database size~$N_2\geq N_r$ with corresponding parameters $\delta t_2$, $\delta\theta_2$ to have another process with~$f_2(t)$ and~$P_2(t)$ which is an approximately time scaled copy of the first process with $\delta t_2 \approx \delta t_1\sqrt{N_2/N_1}$:
\begin{align}
&f_2(t)\approx f_1(t\sqrt{N_2/N_1}), \\ 
&P_2(t)\approx P_1(t\sqrt{N_2/N_1}), \\
&N_2= \left\lceil { N_1 \left( \dfrac{\pi \left\lceil{ \frac{1}{\pi}\sqrt{\frac{N_r}{N_1}}(k_1\pi + \tau) - \frac{\tau}{\pi}  }\right\rceil + \tau}{\pi k_1 + \tau}  \right)^2 }\right\rceil,\label{N2-scaling}
\end{align}
and as can be seen~$N_2/N_r\rightarrow1$ with~$N_r/N_1\rightarrow\infty$.

For a numerical example we chose $N=10^6$ and $\alpha=0.3$, $\tau=0.2$, $k=1$, $\delta t = k\pi+\tau$ as a reference process which gives desired~$V(n)$.
Using~(\ref{N2-scaling}) we find larger database size (
$N_r=10^{18}$), and corresponding integer value of the parameter~$k$ which gives the same~$V(n_G)$ as our reference process ($n_G = \lfloor \pi \sqrt{N}/ (2\delta t)\rfloor$ is the number of steps corresponing to~$T_0$). 
The resulting processes are depicted in Fig.~\ref{fig-scaling}.
Note, the expression~(\ref{k2}) is not satisfied exactly for the process (b) in Fig.~\ref{fig-scaling}. 
For requested~$N_r=10^{18}$ we have~$N_2=1.000000162505052417\times10^{18}$, and from~(\ref{k2}) we have $k_2 = 1063662.0000000002$, however we can see that $k_2 - \lfloor k_2 \rfloor \ll 1$, and for our particular choice we have $k_2 - \lfloor k_2 \rfloor \ll \tau$. 
Processes depicted in Fig.~\ref{fig-scaling} have almost the same survival probability~$P(n_G)\approx 0.27$ and $f(n_G)\approx 0.98$.
At last, the chosen numerical parameters satisfy the condition for saturation dynamics which is discussed in the previous section: $\sqrt{x\delta t}\ll\tau\ll1$ (as well as $0.06\ll0.2\ll1$), and we can see from Fig.~\ref{fig-scaling} that our nonunitary process shows saturation and has an attractor~$\ket{w}$.

\begin{figure}[t]
    \begin{center}
        \includegraphics{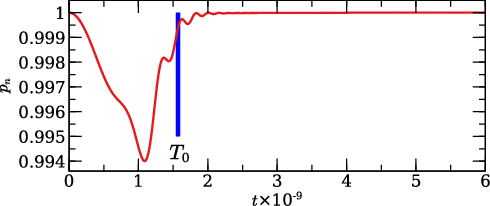}
    \end{center}
    \caption{Success probability of a single step of the computation process as a function of time, with parameters corresponded to~Fig.\ref{fig-scaling}(b). In saturation regime we have~$p_n\rightarrow1$. $T_0\approx1.57\times10^9$.}
    \label{p-n}
\end{figure}

To provide some insight into the the features observed in Fig.~\ref{fig-scaling}, we have plotted the success probabilities $p_n$ for each step $n$ in Fig.~\ref{p-n}. 
We emphasize that at the beginning of the evolution, the success probabilities are relatively far below unity, which indicates that this process cannot be categorized as a standard QZE.
However, in the saturation regime, we obtain $p_n\rightarrow1$, hence the process becomes more Zeno-like.

\section{Robustness against oracle detuning error} 

The continuous time search algorithm requires a high level of accuracy. 
For instance, if we have a detuned oracle~\cite{Grover_with_noise} described by
\begin{equation}
H_o^{(\varepsilon)} = (1+\varepsilon)H_o,
\label{H-o-epsilon}
\end{equation}
then the fidelity of the output state~$f(t) =|\braket{w|U(t)|s}|^2$
 can be low even for $|\varepsilon|\ll1$ (see e.g.\ Ref.~\cite{Grover_with_noise, Novo2018}). 

This section is dedicated to the study of the performance of the nonunitary algorithm under the effect of two kinds of detuning errors.
In both scenarios, we consider the procedure of stopping the evolution (calculation) at time $T_0$ or step $n_G$, for the Hamiltonian and the nonunitary algorithms, respectively, which would yield optimal search performances if the systems were free of errors.
 
\begin{figure}
    \begin{center}
        \includegraphics[scale=1]{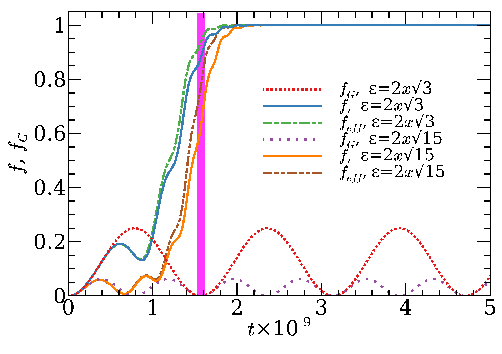}
    \end{center}
    \caption{Exact fidelity~$f$ and approximate fidelity~$f_{\mathrm{eff}}$ calculated using non-hermitian Hamiltonian~(\ref{H-eff}) for nonunitary process, and fidelity~$f_G$ for unitary Grover search as functions of time for two different values of detuning~$\varepsilon$. The vertical solid line corresponds to the time moment when the final measurement is performed.}
    \label{fig-with-errors-Q-a}
\end{figure}

Our first study concerns the static detuning of the oracle Hamiltonian, i.e. when $\varepsilon$ has some fixed unknown value.
It has been shown that stopping the evolution after the Grover time $T_0$ completely fails if the error parameter satisfies a ``resonance condition'' $\varepsilon = \pm 2x\sqrt{4m^2-1},$ for integer $m$ \cite{Novo2018}, and behaves unfavourably in the neighbourhoods of these points.
We carried out numerical simulations of both the original and the nonunitary search algorithm with parameters corresponding to panel (b) in Fig.~\ref{fig-scaling} and using two different values of $\varepsilon$ satisfying the ``resonance condition:'' $\varepsilon = 2x\sqrt3$ and $\varepsilon = 2x\sqrt{15}$.
For comparison, in Fig.~\ref{fig-with-errors-Q-a} we have plotted the fidelity given by the original search Hamiltonian ($f_G$), the target fidelity of our nonunitary algorithm ($f$), as well as the target fidelity~$f_{\mathrm{eff}}$ given by the approximate effective non-hermitian Hamiltonian description.
The latter case corresponds to extending the Hamiltonian in Eq.~(\ref{H-eff}) by an additional term corresponding to the oracle error: $H_{\mathrm{eff}}(t)\rightarrow H_{\mathrm{eff}}(t) - \varepsilon\ket{w}\bra{w}$.
The thick vertical line in Fig.~\ref{fig-with-errors-Q-a} indicates the readout step $n_G$ chosen obliviously to the error parameter $\varepsilon$.
We have $P(n_G)\approx 0.08$, $f(n_G)\approx 0.88$ for $m=1$, and $P(n_G)\approx 0.018$, $f(n_G)\approx 0.63$ for $m=2$.

\begin{figure}
    \begin{center}
        \includegraphics[scale=1]{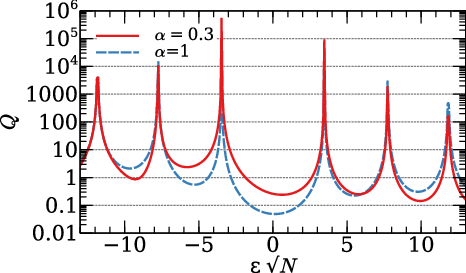}
    \end{center}
    \caption{The $Q$ factor versus oracle detuning~$\varepsilon$ for~$N=10^{18}$, $\delta t=10^6\pi+\tau$, $\tau=0.2$. Peaks correspond to the ``resonance condition''.}
    \label{fig-with-errors-Q-b}
\end{figure}

In order to compare the performance of our nonunitary protocol with continuous Grover search, let us introduce the ``quality factor'' $Q = f(n_G)P(n_G) / f_G(\pi/2x)$. 
As can be seen, $Q>1$ ($Q<1$) corresponds to the case when our proposed nonunitary algorithm is better (worse) than the standard Grover algorithm (we assume that measurements are instantaneous). 
In Fig.~\ref{fig-with-errors-Q-b} we show the dependence of~$Q$ as a function of the systematic oracle error~$\varepsilon$. 
As can be seen, there is a set of intervals of~$\varepsilon$ values where our nonunitary algorithm fares clearly better than the standard continuous search algorithm. 

\begin{figure}[t]
    \begin{center}
        \includegraphics{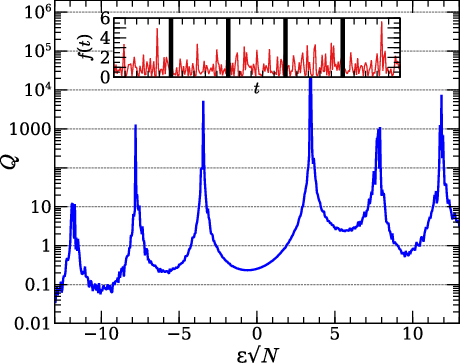}
    \end{center}
    \caption{The $Q$ factor versus strength of the noise~$\varepsilon$ in control for~$N=10^{18}$, $\delta t=10^6\pi+\tau$, $\tau=0.2$, $\alpha=0.3$. In the inset the biased white noise example is shown. Vertical solid lines correspond to the ancilla measurements, and a single realization of the noisy process was made for each value of~$\varepsilon$. Noise function $f(t)$ is modeled by rectangle pulses with duration~$\delta t/50$ and amplitude $-\log(R)$, where $R$ is a uniform random number from interval~$(0,1)$.}
    \label{noise-q}
\end{figure}

Now we turn to the analysis of the time-dependent detuning of the oracle, or a noisy driving of the Hamiltonian.
In particular instead of Eq.~(\ref{ancila-H}) we consider the following Hamiltonian
\begin{equation}
H^{\rm (noisy)}(t)= -\mathcal{I}\otimes\ket{w}\bra{w}-(1 + \varepsilon f(t))\sigma_z\otimes\ket{s}\bra{s},
\end{equation}
where the noise function~$f(t)$ has a mean value of $\overline{f(t)}=1$.
Our numerical results for the $Q$ value are displayed in Fig.~\ref{noise-q} we plot the $Q$ value for the case of noisy control with $f(t)$ modeled as a biased white noise (see inset of Fig.~\ref{noise-q}).
As can be seen, the improvement in robustness is similar to that of constant detuning, shown in Fig.~\ref{fig-with-errors-Q-b}.

\section{Relations to other algorithms using measurement}

While our algorithm resembles the standard quantum Zeno-effect in that frequent measurements are carried out on the dynamical system and the continuation of the evolution is conditioned upon a correct measurement outcome \cite{facchi_quantum_2008}, there are significant differences.
The quantum Zeno effect can be observed when the survival probability at each measurement can be engineered to be arbitrarily close to unity, yielding a process that shares many features with adiabatic processes \cite{burgarth_generalized_2019}.
As a consequence, given a unitary evolution and projective measurements, the resulting Zeno dynamics in a restricted subspace becomes unitary as well.
The quantum Zeno effect in quantum computing has found applications such as error correction \cite{paz-silva_zeno_2012} and algorithm design \cite{rudolph_quantum_2002,hosten_counterfactual_2006,bomb-complexity-lin_et_al}.
Common to these algorithmic applications is that the database registers are measured to induce the Zeno dynamics, while our algorithm uses measurements of the ancilla only.
The quantum circuits studied in Refs.~\cite{rudolph_quantum_2002, bomb-complexity-lin_et_al} operate in the regime where the survival probability at each iteration is kept arbitrarily close to unity, thereby operating in the standard quantum Zeno regime.

The notion of bomb query complexity introduced by Lin et al. \cite{bomb-complexity-lin_et_al} showed a remarkable relation between Zeno dynamics and quantum query complexity, namely that the former is quadratically worse than the latter.
However, since observation can lead to more complex dynamics \cite{burgarth_exponential_2014}, this result appears to be specific to the simple measurement scheme and coupling between the ancilla and the oracle operation. As an example, we have shown here that our algorithm can retain the scaling of the quantum search algorithm.

Closest in spirit to our algorithm is the reported quantum search algorithm based on counterfactual quantum computing \cite{hosten_counterfactual_2006}.
Both approaches start with less than 1 survival probabilities in the initial iterations and quickly converge to a regime where the required measurement outcome is obtained with almost certainty, while exhibiting robustness against certain errors.
The counterfactual search, however, suffers from a conceptual \cite{mitchison_limits_2006} and several technical drawbacks that our approach avoids.
Namely, it requires the implementation of both the oracle and its adjoint, and it might be strongly affected by systematic errors in the oracle.
In addition, no scalable analysis is presented in the paper, therefore it is not clear how the non-unit survival probability scales with the database size.

Finally, while the algorithm combining adiabatic quantum search by measurement employs a rather similar idea of coupling to an ancilla \cite{Farhi-search-by-measurement}, it uses a term $H_o\otimes p$ which is physically very non-trivial.
In contrast, our coupling term is completely oblivious to the marked node.

\section{Discussion and conclusion}
In the case without detuning of the oracle, both the success probability of each step and the target fidelity are approaching~1 (see solid lines in Fig.\ref{fig-scaling} and Fig.\ref{p-n}).
The saturation of the success probability makes the algorithm ideal for applications where verification queries to the classical database are impossible or come at a high cost.
While the algorithm carries strong similarities with the quantum Zeno effect \cite{misra_zenos_1977, exner_open_1985}, and is based on the approach and techniques of Zeno-like dynamics in specific physical systems \cite{Hiromichi,Li2011,Zeno-Polaron-PRB} it employs a time-dependent sequence of projections (via parameter $\theta_n$). The protocol therefore shows similarities also with the proposal of von Neumann to transform quantum states by measurement \cite{von_neumann_mathematical_1955_zeno}, but in our protocol the final state is selected by the oracle Hamiltonian.


The synergy of quantum and classical computation in our algorithm is the result of selective measurements of the ancilla. The classical contribution in the efficiency~$Q$ defined above is determined by the probability~$P\leq1$ (in pure quantum case $P=1$). The quantum/classical ratio of the algorithm can be tuned by properly choosing time intervals~$\delta t$ between subsequent measurements. We note that our algorithm involves only a single quantum process, differing from studies considering the combined application of quantum and classical computers~\cite{Classical-plus-quantum-computation}.

In summary, we have presented a resource-efficient and robust nonunitary modification of the continuous version of Grover's search based on measurements and conditional evolution.
We have shown that this dynamics can be accurately described by a non-hermitian effective Hamiltonian in the Hilbert space~$\mathcal{H}_N$, and determined its scaling properties under ideal conditions.

There are many possible directions for further development of similar protocols. 
For instance, the driven part of Hamiltonian, the parameters~$\delta t$ and~$\delta\theta$ can be made time-dependent (see e.g.\ \cite{Grover-qubit-monitor-Wilczek_2020}). 
We believe that the presented protocol will serve for a better understanding of controllable nonunitary processes and their scaling properties. 
\section*{Acknowledgements}
P.~V.~P. and A.~G. are supported by the National Research,
Development and Innovation Office of Hungary (NKFIH)
Project No. K124351; P.~V.~P.~by NKFIH through the
Projects No. K115624, No. PD120975, No. 2017-1.2.1-NKP-2017-00001; and A.~G.~by MŠMT RVO 14000. P.~V.~P.~and L.~A.~W.~are supported by Grant No. PGC2018-101355-B-I00 funded by MCIN/AEI/10.13039/501100011033 and
by ``ERDF A way of making Europe''. L.~A.~W.~is supported by the Grant No. PID2021-126273NB-I00 funded
by MCIN/AEI/10.13039/501100011033, and the Basque
Government through Grant No. IT1470-22. J.~Q.~Y.~is supported by the National Natural Science Foundation of
China (Grant No. 11934010). P.~V.~P.~thanks Olga Masko for her support.

\appendix*
\section{Condition for saturation behavior of the target fidelity}

We employ the following heuristic analysis to estimate the ranges of parameters as well as the convergence time.
Instead of Eq.~(\ref{H-eff}) we consider the following non-hermitian time-independent Hamiltonian:
\begin{equation} 
h = h_0 - i\gamma\ket{r}\bra{r},
\end{equation}
where $h_0=-x(\ket{w}\bra{r}+\ket{r}\bra{w})$ and $x>0$, $\gamma>0$. 
We can write the non-normalized general solution of Schr\"{o}dinger equation as 
\begin{equation} 
\ket{\psi(t)} = \sum_n e^{-i\epsilon_n t}\braket{\chi_n|\psi(0)}\ket{\varphi_n},
\end{equation} 
where $h\ket{\varphi_n} = \epsilon_n\ket{\varphi_n}$, $h^\dagger\ket{\chi_n} = \lambda_n\ket{\chi_n}$, $\braket{\chi_n|\varphi_m}=\delta_{nm}$, $\epsilon_n=\lambda_n^*$, and $\ket{\psi(0)}=\ket{s}$ is the initial state. 
One can see that in the limit~$\gamma\gg x$ we have $\epsilon_1\approx -ix^2/\gamma \rightarrow 0$, $\epsilon_2\approx -i\gamma$, and thus for $t\gg \gamma^{-1}$ we have $\ket{\psi(t)} \approx e^{-x^2t/\gamma}\braket{\chi_1|\psi(0)}\ket{\varphi_1}$ with $\braket{\chi_1|\psi(0)}\approx -x/\gamma$ and $\ket{\varphi_1}\approx (1; \; -ix/\gamma)^{T}(1+(x/\gamma)^2)^{-1/2}$.

In other words we have a saturation where the state approximates approaches $\ket{w}$ state when $x\ll\gamma$ and $t > x^{-1}$ with target fidelity $|\braket{w|\psi(t)}|^2 \propto x^2 = 1/N$ which corresponds to the classical search efficiency. 
By turning off damping~$\gamma\rightarrow0$ we can achieve a unitary process with quantum efficiency. 
The Hamiltonian in Eq.~(\ref{H-eff}) is non-hermitian and time-dependent and its complete analysis is complicated (see e.g.\  \cite{non-herm-evol-two-level-quant-sys-PhysRevA.42.1467,non-herm-evol-two-level-quant-sys-Bagchi2018}). 
In order to have a general picture of the role of control parameters we collate our simple example with Eq.~(\ref{H-eff}) and approximately set $\gamma\approx \tau^2/\delta t$. 
Thus we can expect saturation regime in the case $\tau \gg \sqrt{x\delta t}$. 
By definition~$\tau$ has a finite value, and moreover, we assume have assumed $\tau\ll1$ whine we deriving Eq.~(\ref{H-eff}), thus we have to choose 
\begin{equation} 
\sqrt{x\delta t}\ll\tau\ll1
\end{equation} 
in order to have saturation property of the computation process.

This requirement also does not contradict our assumption $x\delta t\ll1$ which was made in the main text.
Actually, this assumption has a clear physical meaning: we have to make many measurement cycles during the common Grover time~$T_0$ defined earlier. 
Finally, if $\sqrt{x\delta t}\gg \tau \rightarrow 0$ we do not have a saturation because the process becomes close to unitary.

\bibliographystyle{apsrev4-1}

\begin{thebibliography}{52}%
	\makeatletter
	\providecommand \@ifxundefined [1]{%
		\@ifx{#1\undefined}
	}%
	\providecommand \@ifnum [1]{%
		\ifnum #1\expandafter \@firstoftwo
		\else \expandafter \@secondoftwo
		\fi
	}%
	\providecommand \@ifx [1]{%
		\ifx #1\expandafter \@firstoftwo
		\else \expandafter \@secondoftwo
		\fi
	}%
	\providecommand \natexlab [1]{#1}%
	\providecommand \enquote  [1]{``#1''}%
	\providecommand \bibnamefont  [1]{#1}%
	\providecommand \bibfnamefont [1]{#1}%
	\providecommand \citenamefont [1]{#1}%
	\providecommand \href@noop [0]{\@secondoftwo}%
	\providecommand \href [0]{\begingroup \@sanitize@url \@href}%
	\providecommand \@href[1]{\@@startlink{#1}\@@href}%
	\providecommand \@@href[1]{\endgroup#1\@@endlink}%
	\providecommand \@sanitize@url [0]{\catcode `\\12\catcode `\$12\catcode
		`\&12\catcode `\#12\catcode `\^12\catcode `\_12\catcode `\%12\relax}%
	\providecommand \@@startlink[1]{}%
	\providecommand \@@endlink[0]{}%
	\providecommand \url  [0]{\begingroup\@sanitize@url \@url }%
	\providecommand \@url [1]{\endgroup\@href {#1}{\urlprefix }}%
	\providecommand \urlprefix  [0]{URL }%
	\providecommand \Eprint [0]{\href }%
	\providecommand \doibase [0]{https://doi.org/}%
	\providecommand \selectlanguage [0]{\@gobble}%
	\providecommand \bibinfo  [0]{\@secondoftwo}%
	\providecommand \bibfield  [0]{\@secondoftwo}%
	\providecommand \translation [1]{[#1]}%
	\providecommand \BibitemOpen [0]{}%
	\providecommand \bibitemStop [0]{}%
	\providecommand \bibitemNoStop [0]{.\EOS\space}%
	\providecommand \EOS [0]{\spacefactor3000\relax}%
	\providecommand \BibitemShut  [1]{\csname bibitem#1\endcsname}%
	\let\auto@bib@innerbib\@empty
	\bibitem [{\citenamefont {Facchi}\ and\ \citenamefont
		{Pascazio}(2008{\natexlab{a}})}]{Facchi_2008}%
	\BibitemOpen
	\bibfield  {author} {\bibinfo {author} {\bibfnamefont {P.}~\bibnamefont
			{Facchi}}\ and\ \bibinfo {author} {\bibfnamefont {S.}~\bibnamefont
			{Pascazio}},\ }\bibfield  {title} {\bibinfo {title} {Quantum zeno dynamics:
			mathematical and physical aspects},\ }\href
	{https://doi.org/10.1088/1751-8113/41/49/493001} {\bibfield  {journal}
		{\bibinfo  {journal} {Journal of Physics A: Mathematical and Theoretical}\
		}\textbf {\bibinfo {volume} {41}},\ \bibinfo {pages} {493001} (\bibinfo
		{year} {2008}{\natexlab{a}})}\BibitemShut {NoStop}%
	\bibitem [{\citenamefont {Rudolph}\ and\ \citenamefont
		{Grover}(2002)}]{rudolph_quantum_2002}%
	\BibitemOpen
	\bibfield  {author} {\bibinfo {author} {\bibfnamefont {T.}~\bibnamefont
			{Rudolph}}\ and\ \bibinfo {author} {\bibfnamefont {L.}~\bibnamefont
			{Grover}},\ }\bibfield  {title} {\bibinfo {title} {Quantum searching a
			classical database (or how we learned to stop worrying and love the bomb)},\
	}\href@noop {} {\bibfield  {journal} {\bibinfo  {journal} {ArXiv}\ }\textbf
		{\bibinfo {volume} {quant-ph/0206066v}} (\bibinfo {year} {2002})}\BibitemShut
	{NoStop}%
	\bibitem [{\citenamefont {Childs}\ \emph {et~al.}(2002)\citenamefont {Childs},
		\citenamefont {Deotto}, \citenamefont {Farhi}, \citenamefont {Goldstone},
		\citenamefont {Gutmann},\ and\ \citenamefont
		{Landahl}}]{Farhi-search-by-measurement}%
	\BibitemOpen
	\bibfield  {author} {\bibinfo {author} {\bibfnamefont {A.~M.}\ \bibnamefont
			{Childs}}, \bibinfo {author} {\bibfnamefont {E.}~\bibnamefont {Deotto}},
		\bibinfo {author} {\bibfnamefont {E.}~\bibnamefont {Farhi}}, \bibinfo
		{author} {\bibfnamefont {J.}~\bibnamefont {Goldstone}}, \bibinfo {author}
		{\bibfnamefont {S.}~\bibnamefont {Gutmann}},\ and\ \bibinfo {author}
		{\bibfnamefont {A.~J.}\ \bibnamefont {Landahl}},\ }\bibfield  {title}
	{\bibinfo {title} {Quantum search by measurement},\ }\href
	{https://doi.org/10.1103/PhysRevA.66.032314} {\bibfield  {journal} {\bibinfo
			{journal} {Phys. Rev. A}\ }\textbf {\bibinfo {volume} {66}},\ \bibinfo
		{pages} {032314} (\bibinfo {year} {2002})}\BibitemShut {NoStop}%
	\bibitem [{\citenamefont {Hosten}\ \emph {et~al.}(2006)\citenamefont {Hosten},
		\citenamefont {Rakher}, \citenamefont {Barreiro}, \citenamefont {Peters},\
		and\ \citenamefont {Kwiat}}]{hosten_counterfactual_2006}%
	\BibitemOpen
	\bibfield  {author} {\bibinfo {author} {\bibfnamefont {O.}~\bibnamefont
			{Hosten}}, \bibinfo {author} {\bibfnamefont {M.~T.}\ \bibnamefont {Rakher}},
		\bibinfo {author} {\bibfnamefont {J.~T.}\ \bibnamefont {Barreiro}}, \bibinfo
		{author} {\bibfnamefont {N.~A.}\ \bibnamefont {Peters}},\ and\ \bibinfo
		{author} {\bibfnamefont {P.~G.}\ \bibnamefont {Kwiat}},\ }\bibfield  {title}
	{\bibinfo {title} {Counterfactual quantum computation through quantum
			interrogation},\ }\href {https://doi.org/10.1038/nature04523} {\bibfield
		{journal} {\bibinfo  {journal} {Nature}\ }\textbf {\bibinfo {volume} {439}},\
		\bibinfo {pages} {949} (\bibinfo {year} {2006})}\BibitemShut {NoStop}%
	\bibitem [{\citenamefont {Lin}\ and\ \citenamefont
		{Lin}(2015)}]{bomb-complexity-lin_et_al}%
	\BibitemOpen
	\bibfield  {author} {\bibinfo {author} {\bibfnamefont {C.~Y.-Y.}\
			\bibnamefont {Lin}}\ and\ \bibinfo {author} {\bibfnamefont {H.-H.}\
			\bibnamefont {Lin}},\ }\bibfield  {title} {\bibinfo {title} {{Upper Bounds on
				Quantum Query Complexity Inspired by the Elitzur-Vaidman Bomb Tester}},\ }in\
	\href {https://doi.org/10.4230/LIPIcs.CCC.2015.537} {\emph {\bibinfo
			{booktitle} {30th Conference on Computational Complexity (CCC 2015)}}},\
	\bibinfo {series} {Leibniz International Proceedings in Informatics
		(LIPIcs)}, Vol.~\bibinfo {volume} {33},\ \bibinfo {editor} {edited by\
		\bibinfo {editor} {\bibfnamefont {D.}~\bibnamefont {Zuckerman}}}\ (\bibinfo
	{publisher} {Schloss Dagstuhl--Leibniz-Zentrum fuer Informatik},\ \bibinfo
	{address} {Dagstuhl, Germany},\ \bibinfo {year} {2015})\ pp.\ \bibinfo
	{pages} {537--566}\BibitemShut {NoStop}%
	\bibitem [{\citenamefont {Gily{\'e}n}\ \emph {et~al.}(2019)\citenamefont
		{Gily{\'e}n}, \citenamefont {Su}, \citenamefont {Low},\ and\ \citenamefont
		{Wiebe}}]{gilyen_quantum_2019}%
	\BibitemOpen
	\bibfield  {author} {\bibinfo {author} {\bibfnamefont {A.}~\bibnamefont
			{Gily{\'e}n}}, \bibinfo {author} {\bibfnamefont {Y.}~\bibnamefont {Su}},
		\bibinfo {author} {\bibfnamefont {G.~H.}\ \bibnamefont {Low}},\ and\ \bibinfo
		{author} {\bibfnamefont {N.}~\bibnamefont {Wiebe}},\ }\bibfield  {title}
	{\bibinfo {title} {Quantum singular value transformation and beyond:
			Exponential improvements for quantum matrix arithmetics},\ }in\ \href
	{https://doi.org/10.1145/3313276.3316366} {\emph {\bibinfo {booktitle}
			{Proceedings of the 51st {{Annual ACM SIGACT Symposium}} on {{Theory}} of
				{{Computing}}}}},\ \bibinfo {series and number} {{{STOC}} 2019}\ (\bibinfo
	{publisher} {{Association for Computing Machinery}},\ \bibinfo {address}
	{{Phoenix, AZ, USA}},\ \bibinfo {year} {2019})\ pp.\ \bibinfo {pages}
	{193--204}\BibitemShut {NoStop}%
	\bibitem [{\citenamefont {Knill}\ \emph {et~al.}(2001)\citenamefont {Knill},
		\citenamefont {Laflamme},\ and\ \citenamefont {Milburn}}]{knill_scheme_2001}%
	\BibitemOpen
	\bibfield  {author} {\bibinfo {author} {\bibfnamefont {E.}~\bibnamefont
			{Knill}}, \bibinfo {author} {\bibfnamefont {R.}~\bibnamefont {Laflamme}},\
		and\ \bibinfo {author} {\bibfnamefont {G.~J.}\ \bibnamefont {Milburn}},\
	}\bibfield  {title} {\bibinfo {title} {A scheme for efficient quantum
			computation with linear optics},\ }\href {https://doi.org/10.1038/35051009}
	{\bibfield  {journal} {\bibinfo  {journal} {Nature}\ }\textbf {\bibinfo
			{volume} {409}},\ \bibinfo {pages} {46} (\bibinfo {year} {2001})}\BibitemShut
	{NoStop}%
	\bibitem [{\citenamefont {Terashima}\ and\ \citenamefont
		{Ueda}(2005)}]{non-unitary-gates-2005}%
	\BibitemOpen
	\bibfield  {author} {\bibinfo {author} {\bibfnamefont {H.}~\bibnamefont
			{Terashima}}\ and\ \bibinfo {author} {\bibfnamefont {M.}~\bibnamefont
			{Ueda}},\ }\bibfield  {title} {\bibinfo {title} {Nonunitary quantum
			circuit},\ }\href {https://doi.org/10.1142/S0219749905001456} {\bibfield
		{journal} {\bibinfo  {journal} {International Journal of Quantum
				Information}\ }\textbf {\bibinfo {volume} {03}},\ \bibinfo {pages} {633}
		(\bibinfo {year} {2005})}\BibitemShut {NoStop}%
	\bibitem [{\citenamefont {Kendon}\ and\ \citenamefont
		{Sanders}(2005)}]{non-unitary-QW-1-PhysRevA.71.022307}%
	\BibitemOpen
	\bibfield  {author} {\bibinfo {author} {\bibfnamefont {V.}~\bibnamefont
			{Kendon}}\ and\ \bibinfo {author} {\bibfnamefont {B.~C.}\ \bibnamefont
			{Sanders}},\ }\bibfield  {title} {\bibinfo {title} {Complementarity and
			quantum walks},\ }\href {https://doi.org/10.1103/PhysRevA.71.022307}
	{\bibfield  {journal} {\bibinfo  {journal} {Phys. Rev. A}\ }\textbf {\bibinfo
			{volume} {71}},\ \bibinfo {pages} {022307} (\bibinfo {year}
		{2005})}\BibitemShut {NoStop}%
	\bibitem [{\citenamefont {Kendon}\ and\ \citenamefont
		{Maloyer}(2008)}]{non-unitary-QW-2-KENDON2008187}%
	\BibitemOpen
	\bibfield  {author} {\bibinfo {author} {\bibfnamefont {V.}~\bibnamefont
			{Kendon}}\ and\ \bibinfo {author} {\bibfnamefont {O.}~\bibnamefont
			{Maloyer}},\ }\bibfield  {title} {\bibinfo {title} {Optimal computation with
			non-unitary quantum walks},\ }\href
	{https://doi.org/https://doi.org/10.1016/j.tcs.2007.12.011} {\bibfield
		{journal} {\bibinfo  {journal} {Theoretical Computer Science}\ }\textbf
		{\bibinfo {volume} {394}},\ \bibinfo {pages} {187 } (\bibinfo {year}
		{2008})},\ \bibinfo {note} {from Gödel to Einstein: Computability between
		Logic and Physics}\BibitemShut {NoStop}%
	\bibitem [{\citenamefont {Usher}\ \emph {et~al.}(2017)\citenamefont {Usher},
		\citenamefont {Hoban},\ and\ \citenamefont
		{Browne}}]{non-unitary-compute-1-PhysRevA.96.032321}%
	\BibitemOpen
	\bibfield  {author} {\bibinfo {author} {\bibfnamefont {N.}~\bibnamefont
			{Usher}}, \bibinfo {author} {\bibfnamefont {M.~J.}\ \bibnamefont {Hoban}},\
		and\ \bibinfo {author} {\bibfnamefont {D.~E.}\ \bibnamefont {Browne}},\
	}\bibfield  {title} {\bibinfo {title} {Nonunitary quantum computation in the
			ground space of local hamiltonians},\ }\href
	{https://doi.org/10.1103/PhysRevA.96.032321} {\bibfield  {journal} {\bibinfo
			{journal} {Phys. Rev. A}\ }\textbf {\bibinfo {volume} {96}},\ \bibinfo
		{pages} {032321} (\bibinfo {year} {2017})}\BibitemShut {NoStop}%
	\bibitem [{\citenamefont {Nakazato}\ \emph {et~al.}(2003)\citenamefont
		{Nakazato}, \citenamefont {Takazawa},\ and\ \citenamefont
		{Yuasa}}]{Hiromichi}%
	\BibitemOpen
	\bibfield  {author} {\bibinfo {author} {\bibfnamefont {H.}~\bibnamefont
			{Nakazato}}, \bibinfo {author} {\bibfnamefont {T.}~\bibnamefont {Takazawa}},\
		and\ \bibinfo {author} {\bibfnamefont {K.}~\bibnamefont {Yuasa}},\ }\bibfield
	{title} {\bibinfo {title} {Purification through zeno-like measurements},\
	}\href {https://doi.org/10.1103/PhysRevLett.90.060401} {\bibfield  {journal}
		{\bibinfo  {journal} {Phys. Rev. Lett.}\ }\textbf {\bibinfo {volume} {90}},\
		\bibinfo {pages} {060401} (\bibinfo {year} {2003})}\BibitemShut {NoStop}%
	\bibitem [{\citenamefont {Wu}\ \emph {et~al.}(2004)\citenamefont {Wu},
		\citenamefont {Lidar},\ and\ \citenamefont
		{Schneider}}]{Wu_entanglement_generation}%
	\BibitemOpen
	\bibfield  {author} {\bibinfo {author} {\bibfnamefont {L.-A.}\ \bibnamefont
			{Wu}}, \bibinfo {author} {\bibfnamefont {D.~A.}\ \bibnamefont {Lidar}},\ and\
		\bibinfo {author} {\bibfnamefont {S.}~\bibnamefont {Schneider}},\ }\bibfield
	{title} {\bibinfo {title} {Long-range entanglement generation via frequent
			measurements},\ }\href {https://doi.org/10.1103/PhysRevA.70.032322}
	{\bibfield  {journal} {\bibinfo  {journal} {Phys. Rev. A}\ }\textbf {\bibinfo
			{volume} {70}},\ \bibinfo {pages} {032322} (\bibinfo {year}
		{2004})}\BibitemShut {NoStop}%
	\bibitem [{\citenamefont {Li}\ \emph {et~al.}(2011)\citenamefont {Li},
		\citenamefont {Wu}, \citenamefont {Wang},\ and\ \citenamefont
		{Yang}}]{Li2011}%
	\BibitemOpen
	\bibfield  {author} {\bibinfo {author} {\bibfnamefont {Y.}~\bibnamefont
			{Li}}, \bibinfo {author} {\bibfnamefont {L.-A.}\ \bibnamefont {Wu}}, \bibinfo
		{author} {\bibfnamefont {Y.-D.}\ \bibnamefont {Wang}},\ and\ \bibinfo
		{author} {\bibfnamefont {L.-P.}\ \bibnamefont {Yang}},\ }\bibfield  {title}
	{\bibinfo {title} {Nondeterministic ultrafast ground-state cooling of a
			mechanical resonator},\ }\href {https://doi.org/10.1103/PhysRevB.84.094502}
	{\bibfield  {journal} {\bibinfo  {journal} {Phys. Rev. B}\ }\textbf {\bibinfo
			{volume} {84}},\ \bibinfo {pages} {094502} (\bibinfo {year}
		{2011})}\BibitemShut {NoStop}%
	\bibitem [{\citenamefont {Pyshkin}\ \emph {et~al.}(2016)\citenamefont
		{Pyshkin}, \citenamefont {Luo}, \citenamefont {You},\ and\ \citenamefont
		{Wu}}]{gsc_paper}%
	\BibitemOpen
	\bibfield  {author} {\bibinfo {author} {\bibfnamefont {P.~V.}\ \bibnamefont
			{Pyshkin}}, \bibinfo {author} {\bibfnamefont {D.-W.}\ \bibnamefont {Luo}},
		\bibinfo {author} {\bibfnamefont {J.~Q.}\ \bibnamefont {You}},\ and\ \bibinfo
		{author} {\bibfnamefont {L.-A.}\ \bibnamefont {Wu}},\ }\bibfield  {title}
	{\bibinfo {title} {Ground-state cooling of quantum systems via a one-shot
			measurement},\ }\href {https://doi.org/10.1103/PhysRevA.93.032120} {\bibfield
		{journal} {\bibinfo  {journal} {Phys. Rev. A}\ }\textbf {\bibinfo {volume}
			{93}},\ \bibinfo {pages} {032120} (\bibinfo {year} {2016})}\BibitemShut
	{NoStop}%
	\bibitem [{\citenamefont {Pyshkin}\ \emph {et~al.}(2021)\citenamefont
		{Pyshkin}, \citenamefont {Sherman},\ and\ \citenamefont
		{Wu}}]{Zeno-Polaron-PRB}%
	\BibitemOpen
	\bibfield  {author} {\bibinfo {author} {\bibfnamefont {P.~V.}\ \bibnamefont
			{Pyshkin}}, \bibinfo {author} {\bibfnamefont {E.~Y.}\ \bibnamefont
			{Sherman}},\ and\ \bibinfo {author} {\bibfnamefont {L.-A.}\ \bibnamefont
			{Wu}},\ }\bibfield  {title} {\bibinfo {title} {Polaron formation in a spin
			chain by measurement-induced imaginary zeeman field},\ }\href
	{https://doi.org/10.1103/PhysRevB.104.075136} {\bibfield  {journal} {\bibinfo
			{journal} {Phys. Rev. B}\ }\textbf {\bibinfo {volume} {104}},\ \bibinfo
		{pages} {075136} (\bibinfo {year} {2021})}\BibitemShut {NoStop}%
	\bibitem [{\citenamefont {Luchnikov}\ and\ \citenamefont
		{Filippov}(2017)}]{Filippov2017}%
	\BibitemOpen
	\bibfield  {author} {\bibinfo {author} {\bibfnamefont {I.~A.}\ \bibnamefont
			{Luchnikov}}\ and\ \bibinfo {author} {\bibfnamefont {S.~N.}\ \bibnamefont
			{Filippov}},\ }\bibfield  {title} {\bibinfo {title} {{Quantum evolution in
				the stroboscopic limit of repeated measurements}},\ }\href
	{https://doi.org/10.1103/PhysRevA.95.022113} {\bibfield  {journal} {\bibinfo
			{journal} {Phys. Rev. A}\ }\textbf {\bibinfo {volume} {95}},\ \bibinfo
		{pages} {022113} (\bibinfo {year} {2017})}\BibitemShut {NoStop}%
	\bibitem [{\citenamefont {Streltsov}\ \emph {et~al.}(2011)\citenamefont
		{Streltsov}, \citenamefont {Kampermann},\ and\ \citenamefont
		{Bru\ss{}}}]{Streltsov2011_entanglment_by_measurements}%
	\BibitemOpen
	\bibfield  {author} {\bibinfo {author} {\bibfnamefont {A.}~\bibnamefont
			{Streltsov}}, \bibinfo {author} {\bibfnamefont {H.}~\bibnamefont
			{Kampermann}},\ and\ \bibinfo {author} {\bibfnamefont {D.}~\bibnamefont
			{Bru\ss{}}},\ }\bibfield  {title} {\bibinfo {title} {Linking quantum discord
			to entanglement in a measurement},\ }\href
	{https://doi.org/10.1103/PhysRevLett.106.160401} {\bibfield  {journal}
		{\bibinfo  {journal} {Phys. Rev. Lett.}\ }\textbf {\bibinfo {volume} {106}},\
		\bibinfo {pages} {160401} (\bibinfo {year} {2011})}\BibitemShut {NoStop}%
	\bibitem [{\citenamefont {Gily\'{e}n}\ \emph {et~al.}(2015)\citenamefont
		{Gily\'{e}n}, \citenamefont {Kiss},\ and\ \citenamefont {Jex}}]{Gilyen}%
	\BibitemOpen
	\bibfield  {author} {\bibinfo {author} {\bibfnamefont {A.}~\bibnamefont
			{Gily\'{e}n}}, \bibinfo {author} {\bibfnamefont {T.}~\bibnamefont {Kiss}},\
		and\ \bibinfo {author} {\bibfnamefont {I.}~\bibnamefont {Jex}},\ }\bibfield
	{title} {\bibinfo {title} {Exponential sensitivity and its cost in quantum
			physics},\ }\href@noop {} {\bibfield  {journal} {\bibinfo  {journal} {Sci.
				Rep.}\ }\textbf {\bibinfo {volume} {6}},\ \bibinfo {pages} {20076} (\bibinfo
		{year} {2015})}\BibitemShut {NoStop}%
	\bibitem [{\citenamefont {Coles}\ and\ \citenamefont
		{Piani}(2014)}]{Piani2014_entanglment_by_measurements}%
	\BibitemOpen
	\bibfield  {author} {\bibinfo {author} {\bibfnamefont {P.~J.}\ \bibnamefont
			{Coles}}\ and\ \bibinfo {author} {\bibfnamefont {M.}~\bibnamefont {Piani}},\
	}\bibfield  {title} {\bibinfo {title} {Complementary sequential measurements
			generate entanglement},\ }\href {https://doi.org/10.1103/PhysRevA.89.010302}
	{\bibfield  {journal} {\bibinfo  {journal} {Phys. Rev. A}\ }\textbf {\bibinfo
			{volume} {89}},\ \bibinfo {pages} {010302(R)} (\bibinfo {year}
		{2014})}\BibitemShut {NoStop}%
	\bibitem [{\citenamefont {Torres}\ \emph {et~al.}(2017)\citenamefont {Torres},
		\citenamefont {Bern\'ad}, \citenamefont {Alber}, \citenamefont {K\'alm\'an},\
		and\ \citenamefont {Kiss}}]{Torres2017}%
	\BibitemOpen
	\bibfield  {author} {\bibinfo {author} {\bibfnamefont {J.~M.}\ \bibnamefont
			{Torres}}, \bibinfo {author} {\bibfnamefont {J.~Z.}\ \bibnamefont
			{Bern\'ad}}, \bibinfo {author} {\bibfnamefont {G.}~\bibnamefont {Alber}},
		\bibinfo {author} {\bibfnamefont {O.}~\bibnamefont {K\'alm\'an}},\ and\
		\bibinfo {author} {\bibfnamefont {T.}~\bibnamefont {Kiss}},\ }\bibfield
	{title} {\bibinfo {title} {Measurement-induced chaos and quantum state
			discrimination in an iterated tavis-cummings scheme},\ }\href
	{https://doi.org/10.1103/PhysRevA.95.023828} {\bibfield  {journal} {\bibinfo
			{journal} {Phys. Rev. A}\ }\textbf {\bibinfo {volume} {95}},\ \bibinfo
		{pages} {023828} (\bibinfo {year} {2017})}\BibitemShut {NoStop}%
	\bibitem [{\citenamefont {Pyshkin}\ \emph {et~al.}(2017)\citenamefont
		{Pyshkin}, \citenamefont {Luo}, \citenamefont {You},\ and\ \citenamefont
		{Wu}}]{Pyshkin2017}%
	\BibitemOpen
	\bibfield  {author} {\bibinfo {author} {\bibfnamefont {P.~V.}\ \bibnamefont
			{Pyshkin}}, \bibinfo {author} {\bibfnamefont {D.-W.}\ \bibnamefont {Luo}},
		\bibinfo {author} {\bibfnamefont {J.~Q.}\ \bibnamefont {You}},\ and\ \bibinfo
		{author} {\bibfnamefont {L.-A.}\ \bibnamefont {Wu}},\ }\bibfield  {title}
	{\bibinfo {title} {{Nondeterministic quantum computation via ground state
				cooling and ultrafast Grover algorithm}},\ }\href
	{http://arxiv.org/abs/1704.01467} {\bibfield  {journal} {\bibinfo  {journal}
			{ArXiv}\ } (\bibinfo {year} {2017})},\ \Eprint
	{https://arxiv.org/abs/1704.01467} {arXiv:1704.01467} \BibitemShut {NoStop}%
	\bibitem [{\citenamefont {Grimaudo}\ \emph {et~al.}(2020)\citenamefont
		{Grimaudo}, \citenamefont {Messina}, \citenamefont {Sergi}, \citenamefont
		{Vitanov},\ and\ \citenamefont
		{Filippov}}]{two-qubit-entangl-measurements-non-herm}%
	\BibitemOpen
	\bibfield  {author} {\bibinfo {author} {\bibfnamefont {R.}~\bibnamefont
			{Grimaudo}}, \bibinfo {author} {\bibfnamefont {A.}~\bibnamefont {Messina}},
		\bibinfo {author} {\bibfnamefont {A.}~\bibnamefont {Sergi}}, \bibinfo
		{author} {\bibfnamefont {N.~V.}\ \bibnamefont {Vitanov}},\ and\ \bibinfo
		{author} {\bibfnamefont {S.~N.}\ \bibnamefont {Filippov}},\ }\bibfield
	{title} {\bibinfo {title} {Two-qubit entanglement generation through
			non-hermitian hamiltonians induced by repeated measurements on an ancilla},\
	}\href {https://www.mdpi.com/1099-4300/22/10/1184} {\bibfield  {journal}
		{\bibinfo  {journal} {Entropy}\ }\textbf {\bibinfo {volume} {22}},\ \bibinfo
		{pages} {1184} (\bibinfo {year} {2020})}\BibitemShut {NoStop}%
	\bibitem [{\citenamefont {Amin}\ \emph {et~al.}(2008)\citenamefont {Amin},
		\citenamefont {Love},\ and\ \citenamefont {Truncik}}]{Amin2008}%
	\BibitemOpen
	\bibfield  {author} {\bibinfo {author} {\bibfnamefont {M.~H.~S.}\
			\bibnamefont {Amin}}, \bibinfo {author} {\bibfnamefont {P.~J.}\ \bibnamefont
			{Love}},\ and\ \bibinfo {author} {\bibfnamefont {C.~J.~S.}\ \bibnamefont
			{Truncik}},\ }\bibfield  {title} {\bibinfo {title} {Thermally assisted
			adiabatic quantum computation},\ }\href
	{https://doi.org/10.1103/PhysRevLett.100.060503} {\bibfield  {journal}
		{\bibinfo  {journal} {Phys. Rev. Lett.}\ }\textbf {\bibinfo {volume} {100}},\
		\bibinfo {pages} {060503} (\bibinfo {year} {2008})}\BibitemShut {NoStop}%
	\bibitem [{\citenamefont {Luo}\ \emph {et~al.}(2015)\citenamefont {Luo},
		\citenamefont {Pyshkin}, \citenamefont {Lam}, \citenamefont {Yu},
		\citenamefont {Lin}, \citenamefont {You},\ and\ \citenamefont
		{Wu}}]{Luo2015}%
	\BibitemOpen
	\bibfield  {author} {\bibinfo {author} {\bibfnamefont {D.-W.}\ \bibnamefont
			{Luo}}, \bibinfo {author} {\bibfnamefont {P.~V.}\ \bibnamefont {Pyshkin}},
		\bibinfo {author} {\bibfnamefont {C.-H.}\ \bibnamefont {Lam}}, \bibinfo
		{author} {\bibfnamefont {T.}~\bibnamefont {Yu}}, \bibinfo {author}
		{\bibfnamefont {H.-Q.}\ \bibnamefont {Lin}}, \bibinfo {author} {\bibfnamefont
			{J.~Q.}\ \bibnamefont {You}},\ and\ \bibinfo {author} {\bibfnamefont {L.-A.}\
			\bibnamefont {Wu}},\ }\bibfield  {title} {\bibinfo {title} {Dynamical
			invariants in a non-markovian quantum-state-diffusion equation},\ }\href
	{https://doi.org/10.1103/PhysRevA.92.062127} {\bibfield  {journal} {\bibinfo
			{journal} {Phys. Rev. A}\ }\textbf {\bibinfo {volume} {92}},\ \bibinfo
		{pages} {062127} (\bibinfo {year} {2015})}\BibitemShut {NoStop}%
	\bibitem [{\citenamefont {Novo}\ \emph {et~al.}(2018)\citenamefont {Novo},
		\citenamefont {Chakraborty}, \citenamefont {Mohseni},\ and\ \citenamefont
		{Omar}}]{Novo2018}%
	\BibitemOpen
	\bibfield  {author} {\bibinfo {author} {\bibfnamefont {L.}~\bibnamefont
			{Novo}}, \bibinfo {author} {\bibfnamefont {S.}~\bibnamefont {Chakraborty}},
		\bibinfo {author} {\bibfnamefont {M.}~\bibnamefont {Mohseni}},\ and\ \bibinfo
		{author} {\bibfnamefont {Y.}~\bibnamefont {Omar}},\ }\bibfield  {title}
	{\bibinfo {title} {Environment-assisted analog quantum search},\ }\href
	{https://doi.org/10.1103/PhysRevA.98.022316} {\bibfield  {journal} {\bibinfo
			{journal} {Phys. Rev. A}\ }\textbf {\bibinfo {volume} {98}},\ \bibinfo
		{pages} {022316} (\bibinfo {year} {2018})}\BibitemShut {NoStop}%
	\bibitem [{\citenamefont {Farhi}\ and\ \citenamefont
		{Gutmann}(1998)}]{H-oracle-first-Farhi1998}%
	\BibitemOpen
	\bibfield  {author} {\bibinfo {author} {\bibfnamefont {E.}~\bibnamefont
			{Farhi}}\ and\ \bibinfo {author} {\bibfnamefont {S.}~\bibnamefont
			{Gutmann}},\ }\bibfield  {title} {\bibinfo {title} {Analog analogue of a
			digital quantum computation},\ }\href
	{https://doi.org/10.1103/physreva.57.2403} {\bibfield  {journal} {\bibinfo
			{journal} {Phys. Rev. A}\ }\textbf {\bibinfo {volume} {57}},\ \bibinfo
		{pages} {2403} (\bibinfo {year} {1998})}\BibitemShut {NoStop}%
	\bibitem [{\citenamefont {Wu}\ and\ \citenamefont
		{Byrd}(2008)}]{self-protected-algos}%
	\BibitemOpen
	\bibfield  {author} {\bibinfo {author} {\bibfnamefont {L.-A.}\ \bibnamefont
			{Wu}}\ and\ \bibinfo {author} {\bibfnamefont {M.~S.}\ \bibnamefont {Byrd}},\
	}\bibfield  {title} {\bibinfo {title} {Self-protected quantum algorithms
			based on quantum state tomography},\ }\href@noop {} {\bibfield  {journal}
		{\bibinfo  {journal} {Quantum Information Processing}\ }\textbf {\bibinfo
			{volume} {8}},\ \bibinfo {pages} {1} (\bibinfo {year} {2008})}\BibitemShut
	{NoStop}%
	\bibitem [{\citenamefont {Lidar}\ and\ \citenamefont
		{Wu}(2003)}]{PRA-recoupling-and-decoupling}%
	\BibitemOpen
	\bibfield  {author} {\bibinfo {author} {\bibfnamefont {D.~A.}\ \bibnamefont
			{Lidar}}\ and\ \bibinfo {author} {\bibfnamefont {L.-A.}\ \bibnamefont {Wu}},\
	}\bibfield  {title} {\bibinfo {title} {Encoded recoupling and decoupling: An
			alternative to quantum error-correcting codes applied to trapped-ion quantum
			computation},\ }\href {https://doi.org/10.1103/PhysRevA.67.032313} {\bibfield
		{journal} {\bibinfo  {journal} {Phys. Rev. A}\ }\textbf {\bibinfo {volume}
			{67}},\ \bibinfo {pages} {032313} (\bibinfo {year} {2003})}\BibitemShut
	{NoStop}%
	\bibitem [{\citenamefont {{Farhi}}\ \emph {et~al.}(2000)\citenamefont
		{{Farhi}}, \citenamefont {{Goldstone}}, \citenamefont {{Gutmann}},\ and\
		\citenamefont {{Sipser}}}]{Adiabatic_qc_1}%
	\BibitemOpen
	\bibfield  {author} {\bibinfo {author} {\bibfnamefont {E.}~\bibnamefont
			{{Farhi}}}, \bibinfo {author} {\bibfnamefont {J.}~\bibnamefont
			{{Goldstone}}}, \bibinfo {author} {\bibfnamefont {S.}~\bibnamefont
			{{Gutmann}}},\ and\ \bibinfo {author} {\bibfnamefont {M.}~\bibnamefont
			{{Sipser}}},\ }\bibfield  {title} {\bibinfo {title} {{Quantum Computation by
				Adiabatic Evolution}},\ }\href@noop {} {\bibfield  {journal} {\bibinfo
			{journal} {eprint arXiv:quant-ph/0001106}\ } (\bibinfo {year} {2000})},\
	\Eprint {https://arxiv.org/abs/quant-ph/0001106} {quant-ph/0001106}
	\BibitemShut {NoStop}%
	\bibitem [{\citenamefont {Roland}\ and\ \citenamefont
		{Cerf}(2002)}]{Adiabatic_qc_2}%
	\BibitemOpen
	\bibfield  {author} {\bibinfo {author} {\bibfnamefont {J.}~\bibnamefont
			{Roland}}\ and\ \bibinfo {author} {\bibfnamefont {N.~J.}\ \bibnamefont
			{Cerf}},\ }\bibfield  {title} {\bibinfo {title} {Quantum search by local
			adiabatic evolution},\ }\href {https://doi.org/10.1103/PhysRevA.65.042308}
	{\bibfield  {journal} {\bibinfo  {journal} {Phys. Rev. A}\ }\textbf {\bibinfo
			{volume} {65}},\ \bibinfo {pages} {042308} (\bibinfo {year}
		{2002})}\BibitemShut {NoStop}%
	\bibitem [{\citenamefont {P\'erez}\ and\ \citenamefont
		{Romanelli}(2007)}]{non-adiab-search-1}%
	\BibitemOpen
	\bibfield  {author} {\bibinfo {author} {\bibfnamefont {A.}~\bibnamefont
			{P\'erez}}\ and\ \bibinfo {author} {\bibfnamefont {A.}~\bibnamefont
			{Romanelli}},\ }\bibfield  {title} {\bibinfo {title} {Nonadiabatic quantum
			search algorithms},\ }\href {https://doi.org/10.1103/PhysRevA.76.052318}
	{\bibfield  {journal} {\bibinfo  {journal} {Phys. Rev. A}\ }\textbf {\bibinfo
			{volume} {76}},\ \bibinfo {pages} {052318} (\bibinfo {year}
		{2007})}\BibitemShut {NoStop}%
	\bibitem [{\citenamefont {Li}\ \emph {et~al.}(2018)\citenamefont {Li},
		\citenamefont {Bao}, \citenamefont {Li}, \citenamefont {liang Huang},
		\citenamefont {Zhang},\ and\ \citenamefont {Fu}}]{non-adiab-search-2}%
	\BibitemOpen
	\bibfield  {author} {\bibinfo {author} {\bibfnamefont {F.-G.}\ \bibnamefont
			{Li}}, \bibinfo {author} {\bibfnamefont {W.-S.}\ \bibnamefont {Bao}},
		\bibinfo {author} {\bibfnamefont {T.}~\bibnamefont {Li}}, \bibinfo {author}
		{\bibfnamefont {H.}~\bibnamefont {liang Huang}}, \bibinfo {author}
		{\bibfnamefont {S.}~\bibnamefont {Zhang}},\ and\ \bibinfo {author}
		{\bibfnamefont {X.-Q.}\ \bibnamefont {Fu}},\ }\bibfield  {title} {\bibinfo
		{title} {Nonadiabatic quantum search algorithm with analytical success
			rate},\ }\href {https://doi.org/10.1007/s10773-018-3986-x} {\bibfield
		{journal} {\bibinfo  {journal} {International Journal of Theoretical
				Physics}\ }\textbf {\bibinfo {volume} {58}},\ \bibinfo {pages} {939}
		(\bibinfo {year} {2018})}\BibitemShut {NoStop}%
	\bibitem [{\citenamefont {Salerno}(1995)}]{Complete-graph-hopping-Salerno1995}%
	\BibitemOpen
	\bibfield  {author} {\bibinfo {author} {\bibfnamefont {M.}~\bibnamefont
			{Salerno}},\ }\bibfield  {title} {\bibinfo {title} {The hubbard model on a
			complete graph: exact analytical results},\ }\href
	{https://doi.org/10.1007/s002570050064} {\bibfield  {journal} {\bibinfo
			{journal} {Zeitschrift für Physik B Condensed Matter}\ }\textbf {\bibinfo
			{volume} {99}},\ \bibinfo {pages} {469} (\bibinfo {year} {1995})}\BibitemShut
	{NoStop}%
	\bibitem [{\citenamefont {Jacko}\ \emph {et~al.}(2017)\citenamefont {Jacko},
		\citenamefont {Khosla}, \citenamefont {Merino},\ and\ \citenamefont
		{Powell}}]{spin-dependent-hopping-PhysRevB.95.155120}%
	\BibitemOpen
	\bibfield  {author} {\bibinfo {author} {\bibfnamefont {A.~C.}\ \bibnamefont
			{Jacko}}, \bibinfo {author} {\bibfnamefont {A.~L.}\ \bibnamefont {Khosla}},
		\bibinfo {author} {\bibfnamefont {J.}~\bibnamefont {Merino}},\ and\ \bibinfo
		{author} {\bibfnamefont {B.~J.}\ \bibnamefont {Powell}},\ }\bibfield  {title}
	{\bibinfo {title} {Spin-orbit coupling in
			${\mathrm{mo}}_{3}{\mathrm{s}}_{7}{(\mathrm{dmit})}_{3}$},\ }\href
	{https://doi.org/10.1103/PhysRevB.95.155120} {\bibfield  {journal} {\bibinfo
			{journal} {Phys. Rev. B}\ }\textbf {\bibinfo {volume} {95}},\ \bibinfo
		{pages} {155120} (\bibinfo {year} {2017})}\BibitemShut {NoStop}%
	\bibitem [{\citenamefont {Lloyd}(1996)}]{Trotterization-Lloyd1996}%
	\BibitemOpen
	\bibfield  {author} {\bibinfo {author} {\bibfnamefont {S.}~\bibnamefont
			{Lloyd}},\ }\bibfield  {title} {\bibinfo {title} {Universal quantum
			simulators},\ }\href {https://doi.org/10.1126/science.273.5278.1073}
	{\bibfield  {journal} {\bibinfo  {journal} {Science}\ }\textbf {\bibinfo
			{volume} {273}},\ \bibinfo {pages} {1073} (\bibinfo {year}
		{1996})}\BibitemShut {NoStop}%
	\bibitem [{\citenamefont {Wu}\ \emph {et~al.}(2002)\citenamefont {Wu},
		\citenamefont {Byrd},\ and\ \citenamefont {Lidar}}]{Wu2002-trotterization}%
	\BibitemOpen
	\bibfield  {author} {\bibinfo {author} {\bibfnamefont {L.-A.}\ \bibnamefont
			{Wu}}, \bibinfo {author} {\bibfnamefont {M.~S.}\ \bibnamefont {Byrd}},\ and\
		\bibinfo {author} {\bibfnamefont {D.~A.}\ \bibnamefont {Lidar}},\ }\bibfield
	{title} {\bibinfo {title} {Polynomial-time simulation of pairing models on a
			quantum computer},\ }\href {https://doi.org/10.1103/PhysRevLett.89.057904}
	{\bibfield  {journal} {\bibinfo  {journal} {Physical Review Letters}\
		}\textbf {\bibinfo {volume} {89}},\ \bibinfo {pages} {057904} (\bibinfo
		{year} {2002})}\BibitemShut {NoStop}%
	\bibitem [{\citenamefont {Smith}\ \emph {et~al.}(2019)\citenamefont {Smith},
		\citenamefont {Kim}, \citenamefont {Pollmann},\ and\ \citenamefont
		{Knolle}}]{Trotterization-Smith2019}%
	\BibitemOpen
	\bibfield  {author} {\bibinfo {author} {\bibfnamefont {A.}~\bibnamefont
			{Smith}}, \bibinfo {author} {\bibfnamefont {M.~S.}\ \bibnamefont {Kim}},
		\bibinfo {author} {\bibfnamefont {F.}~\bibnamefont {Pollmann}},\ and\
		\bibinfo {author} {\bibfnamefont {J.}~\bibnamefont {Knolle}},\ }\bibfield
	{title} {\bibinfo {title} {Simulating quantum many-body dynamics on a current
			digital quantum computer},\ }\href
	{https://doi.org/10.1038/s41534-019-0217-0} {\bibfield  {journal} {\bibinfo
			{journal} {npj Quantum Information}\ }\textbf {\bibinfo {volume} {5}},\
		\bibinfo {pages} {106} (\bibinfo {year} {2019})}\BibitemShut {NoStop}%
	\bibitem [{\citenamefont {Tacchino}\ \emph {et~al.}(2020)\citenamefont
		{Tacchino}, \citenamefont {Chiesa}, \citenamefont {Carretta},\ and\
		\citenamefont {Gerace}}]{Trotterization-Tacchino2020}%
	\BibitemOpen
	\bibfield  {author} {\bibinfo {author} {\bibfnamefont {F.}~\bibnamefont
			{Tacchino}}, \bibinfo {author} {\bibfnamefont {A.}~\bibnamefont {Chiesa}},
		\bibinfo {author} {\bibfnamefont {S.}~\bibnamefont {Carretta}},\ and\
		\bibinfo {author} {\bibfnamefont {D.}~\bibnamefont {Gerace}},\ }\bibfield
	{title} {\bibinfo {title} {Quantum computers as universal quantum simulators:
			State‐of‐the‐art and perspectives},\ }\href
	{https://doi.org/10.1002/qute.201900052} {\bibfield  {journal} {\bibinfo
			{journal} {Advanced Quantum Technologies}\ }\textbf {\bibinfo {volume} {3}},\
		\bibinfo {pages} {1900052} (\bibinfo {year} {2020})}\BibitemShut {NoStop}%
	\bibitem [{\citenamefont {Shenvi}\ \emph {et~al.}(2003)\citenamefont {Shenvi},
		\citenamefont {Brown},\ and\ \citenamefont {Whaley}}]{Grover_with_noise}%
	\BibitemOpen
	\bibfield  {author} {\bibinfo {author} {\bibfnamefont {N.}~\bibnamefont
			{Shenvi}}, \bibinfo {author} {\bibfnamefont {K.~R.}\ \bibnamefont {Brown}},\
		and\ \bibinfo {author} {\bibfnamefont {K.~B.}\ \bibnamefont {Whaley}},\
	}\bibfield  {title} {\bibinfo {title} {Effects of a random noisy oracle on
			search algorithm complexity},\ }\href
	{https://doi.org/10.1103/PhysRevA.68.052313} {\bibfield  {journal} {\bibinfo
			{journal} {Phys. Rev. A}\ }\textbf {\bibinfo {volume} {68}},\ \bibinfo
		{pages} {052313} (\bibinfo {year} {2003})}\BibitemShut {NoStop}%
	\bibitem [{\citenamefont {Facchi}\ and\ \citenamefont
		{Pascazio}(2008{\natexlab{b}})}]{facchi_quantum_2008}%
	\BibitemOpen
	\bibfield  {author} {\bibinfo {author} {\bibfnamefont {P.}~\bibnamefont
			{Facchi}}\ and\ \bibinfo {author} {\bibfnamefont {S.}~\bibnamefont
			{Pascazio}},\ }\bibfield  {title} {\bibinfo {title} {Quantum {{Zeno}}
			dynamics: Mathematical and physical aspects},\ }\href
	{https://doi.org/10.1088/1751-8113/41/49/493001} {\bibfield  {journal}
		{\bibinfo  {journal} {J. Phys. A: Math. Theor.}\ }\textbf {\bibinfo {volume}
			{41}},\ \bibinfo {pages} {493001} (\bibinfo {year}
		{2008}{\natexlab{b}})}\BibitemShut {NoStop}%
	\bibitem [{\citenamefont {Burgarth}\ \emph {et~al.}(2019)\citenamefont
		{Burgarth}, \citenamefont {Facchi}, \citenamefont {Nakazato}, \citenamefont
		{Pascazio},\ and\ \citenamefont {Yuasa}}]{burgarth_generalized_2019}%
	\BibitemOpen
	\bibfield  {author} {\bibinfo {author} {\bibfnamefont {D.}~\bibnamefont
			{Burgarth}}, \bibinfo {author} {\bibfnamefont {P.}~\bibnamefont {Facchi}},
		\bibinfo {author} {\bibfnamefont {H.}~\bibnamefont {Nakazato}}, \bibinfo
		{author} {\bibfnamefont {S.}~\bibnamefont {Pascazio}},\ and\ \bibinfo
		{author} {\bibfnamefont {K.}~\bibnamefont {Yuasa}},\ }\bibfield  {title}
	{\bibinfo {title} {Generalized {{Adiabatic Theorem}} and {{Strong-Coupling
					Limits}}},\ }\href {https://doi.org/10.22331/q-2019-06-12-152} {\bibfield
		{journal} {\bibinfo  {journal} {Quantum}\ }\textbf {\bibinfo {volume} {3}},\
		\bibinfo {pages} {152} (\bibinfo {year} {2019})}\BibitemShut {NoStop}%
	\bibitem [{\citenamefont {{Paz-Silva}}\ \emph {et~al.}(2012)\citenamefont
		{{Paz-Silva}}, \citenamefont {Rezakhani}, \citenamefont {Dominy},\ and\
		\citenamefont {Lidar}}]{paz-silva_zeno_2012}%
	\BibitemOpen
	\bibfield  {author} {\bibinfo {author} {\bibfnamefont {G.~A.}\ \bibnamefont
			{{Paz-Silva}}}, \bibinfo {author} {\bibfnamefont {A.~T.}\ \bibnamefont
			{Rezakhani}}, \bibinfo {author} {\bibfnamefont {J.~M.}\ \bibnamefont
			{Dominy}},\ and\ \bibinfo {author} {\bibfnamefont {D.~A.}\ \bibnamefont
			{Lidar}},\ }\bibfield  {title} {\bibinfo {title} {Zeno {{Effect}} for
			{{Quantum Computation}} and {{Control}}},\ }\href
	{https://doi.org/10.1103/PhysRevLett.108.080501} {\bibfield  {journal}
		{\bibinfo  {journal} {Phys. Rev. Lett.}\ }\textbf {\bibinfo {volume} {108}},\
		\bibinfo {pages} {080501} (\bibinfo {year} {2012})}\BibitemShut {NoStop}%
	\bibitem [{\citenamefont {Burgarth}\ \emph {et~al.}(2014)\citenamefont
		{Burgarth}, \citenamefont {Facchi}, \citenamefont {Giovannetti},
		\citenamefont {Nakazato}, \citenamefont {Pascazio},\ and\ \citenamefont
		{Yuasa}}]{burgarth_exponential_2014}%
	\BibitemOpen
	\bibfield  {author} {\bibinfo {author} {\bibfnamefont {D.~K.}\ \bibnamefont
			{Burgarth}}, \bibinfo {author} {\bibfnamefont {P.}~\bibnamefont {Facchi}},
		\bibinfo {author} {\bibfnamefont {V.}~\bibnamefont {Giovannetti}}, \bibinfo
		{author} {\bibfnamefont {H.}~\bibnamefont {Nakazato}}, \bibinfo {author}
		{\bibfnamefont {S.}~\bibnamefont {Pascazio}},\ and\ \bibinfo {author}
		{\bibfnamefont {K.}~\bibnamefont {Yuasa}},\ }\bibfield  {title} {\bibinfo
		{title} {Exponential rise of dynamical complexity in quantum computing
			through projections},\ }\href {https://doi.org/10.1038/ncomms6173} {\bibfield
		{journal} {\bibinfo  {journal} {Nat Commun}\ }\textbf {\bibinfo {volume}
			{5}},\ \bibinfo {pages} {5173} (\bibinfo {year} {2014})}\BibitemShut
	{NoStop}%
	\bibitem [{\citenamefont {Mitchison}\ and\ \citenamefont
		{Jozsa}(2006)}]{mitchison_limits_2006}%
	\BibitemOpen
	\bibfield  {author} {\bibinfo {author} {\bibfnamefont {G.}~\bibnamefont
			{Mitchison}}\ and\ \bibinfo {author} {\bibfnamefont {R.}~\bibnamefont
			{Jozsa}},\ }\bibfield  {title} {\bibinfo {title} {The limits of
			counterfactual computation},\ }\href@noop {} {\bibfield  {journal} {\bibinfo
			{journal} {ArXiv}\ }\textbf {\bibinfo {volume} {quant-ph/0606092v3}}
		(\bibinfo {year} {2006})}\BibitemShut {NoStop}%
	\bibitem [{\citenamefont {Misra}\ and\ \citenamefont
		{Sudarshan}(1977)}]{misra_zenos_1977}%
	\BibitemOpen
	\bibfield  {author} {\bibinfo {author} {\bibfnamefont {B.}~\bibnamefont
			{Misra}}\ and\ \bibinfo {author} {\bibfnamefont {E.~C.~G.}\ \bibnamefont
			{Sudarshan}},\ }\bibfield  {title} {\bibinfo {title} {The {{Zeno}}'s paradox
			in quantum theory},\ }\href {https://doi.org/10.1063/1.523304} {\bibfield
		{journal} {\bibinfo  {journal} {Journal of Mathematical Physics}\ }\textbf
		{\bibinfo {volume} {18}},\ \bibinfo {pages} {756} (\bibinfo {year}
		{1977})}\BibitemShut {NoStop}%
	\bibitem [{\citenamefont {Exner}(1985)}]{exner_open_1985}%
	\BibitemOpen
	\bibfield  {author} {\bibinfo {author} {\bibfnamefont {P.}~\bibnamefont
			{Exner}},\ }\href {https://doi.org/10.1007/978-94-009-5207-2} {\emph
		{\bibinfo {title} {Open {{Quantum Systems}} and {{Feynman Integrals}}}}},\
	Fundamental {{Theories}} of {{Physics}}\ (\bibinfo  {publisher} {{Springer
			Netherlands}},\ \bibinfo {year} {1985})\BibitemShut {NoStop}%
	\bibitem [{\citenamefont {{von
				Neumann}}(1955)}]{von_neumann_mathematical_1955_zeno}%
	\BibitemOpen
	\bibfield  {author} {\bibinfo {author} {\bibfnamefont {J.}~\bibnamefont {{von
					Neumann}}},\ }\bibinfo {title} {Mathematical {{Foundations}} of {{Quantum
				Mechanics}}}\ (\bibinfo  {publisher} {{Princeton Univ Press}},\ \bibinfo
	{address} {Princeton},\ \bibinfo {year} {1955})\ pp.\ \bibinfo {pages}
	{365--367}\BibitemShut {NoStop}%
	\bibitem [{\citenamefont {Bravyi}\ \emph {et~al.}(2016)\citenamefont {Bravyi},
		\citenamefont {Smith},\ and\ \citenamefont
		{Smolin}}]{Classical-plus-quantum-computation}%
	\BibitemOpen
	\bibfield  {author} {\bibinfo {author} {\bibfnamefont {S.}~\bibnamefont
			{Bravyi}}, \bibinfo {author} {\bibfnamefont {G.}~\bibnamefont {Smith}},\ and\
		\bibinfo {author} {\bibfnamefont {J.~A.}\ \bibnamefont {Smolin}},\ }\bibfield
	{title} {\bibinfo {title} {Trading classical and quantum computational
			resources},\ }\href {https://doi.org/10.1103/PhysRevX.6.021043} {\bibfield
		{journal} {\bibinfo  {journal} {Phys. Rev. X}\ }\textbf {\bibinfo {volume}
			{6}},\ \bibinfo {pages} {021043} (\bibinfo {year} {2016})}\BibitemShut
	{NoStop}%
	\bibitem [{\citenamefont {Wilczek}\ \emph {et~al.}(2020)\citenamefont
		{Wilczek}, \citenamefont {Hu},\ and\ \citenamefont
		{Wu}}]{Grover-qubit-monitor-Wilczek_2020}%
	\BibitemOpen
	\bibfield  {author} {\bibinfo {author} {\bibfnamefont {F.}~\bibnamefont
			{Wilczek}}, \bibinfo {author} {\bibfnamefont {H.-Y.}\ \bibnamefont {Hu}},\
		and\ \bibinfo {author} {\bibfnamefont {B.}~\bibnamefont {Wu}},\ }\bibfield
	{title} {\bibinfo {title} {Resonant quantum search with monitor qubits},\
	}\href {https://doi.org/10.1088/0256-307x/37/5/050304} {\bibfield  {journal}
		{\bibinfo  {journal} {Chinese Physics Letters}\ }\textbf {\bibinfo {volume}
			{37}},\ \bibinfo {pages} {050304} (\bibinfo {year} {2020})}\BibitemShut
	{NoStop}%
	\bibitem [{\citenamefont {Dattoli}\ \emph {et~al.}(1990)\citenamefont
		{Dattoli}, \citenamefont {Torre},\ and\ \citenamefont
		{Mignani}}]{non-herm-evol-two-level-quant-sys-PhysRevA.42.1467}%
	\BibitemOpen
	\bibfield  {author} {\bibinfo {author} {\bibfnamefont {G.}~\bibnamefont
			{Dattoli}}, \bibinfo {author} {\bibfnamefont {A.}~\bibnamefont {Torre}},\
		and\ \bibinfo {author} {\bibfnamefont {R.}~\bibnamefont {Mignani}},\
	}\bibfield  {title} {\bibinfo {title} {Non-hermitian evolution of two-level
			quantum systems},\ }\href {https://doi.org/10.1103/PhysRevA.42.1467}
	{\bibfield  {journal} {\bibinfo  {journal} {Phys. Rev. A}\ }\textbf {\bibinfo
			{volume} {42}},\ \bibinfo {pages} {1467} (\bibinfo {year}
		{1990})}\BibitemShut {NoStop}%
	\bibitem [{\citenamefont
		{Bagchi}(2018)}]{non-herm-evol-two-level-quant-sys-Bagchi2018}%
	\BibitemOpen
	\bibfield  {author} {\bibinfo {author} {\bibfnamefont {B.}~\bibnamefont
			{Bagchi}},\ }\bibfield  {title} {\bibinfo {title} {Evolution operator for
			time-dependent non-hermitian hami ltonians},\ }\href
	{https://doi.org/10.31526/lhep.3.2018.02} {\bibfield  {journal} {\bibinfo
			{journal} {Letters in High Energy Physics}\ }\textbf {\bibinfo {volume}
			{1}},\ \bibinfo {pages} {4} (\bibinfo {year} {2018})}\BibitemShut {NoStop}%
\end{thebibliography}

%

\end{document}